\documentclass[apjl]{emulateapj}
\usepackage{color}

\usepackage{graphicx}

\usepackage{rotating}

\begin{document}

\title{On the Mass Function, Multiplicity, and Origins of Wide-Orbit Giant Planets}

\color{black}
\shorttitle{The Mass Function of Wide Orbit Giant Planets}
\shortauthors{Wagner, Apai, \& Kratter}
\author{Kevin Wagner\altaffilmark{1,2,3,}$^{\star}$, D\'aniel Apai\altaffilmark{1,3,4}, \& Kaitlin M. Kratter\altaffilmark{1}}


\altaffiltext{1}{Steward Observatory, University of Arizona}

\altaffiltext{2}{National Science Foundation Graduate Research Fellow}
\altaffiltext{3}{NASA NExSS \textit{Earths in Other Solar Systems} Team}
\altaffiltext{4}{Lunar and Planetary Laboratory, University of Arizona}
\altaffiltext{$\star$}{Correspondence to: kwagner@as.arizona.edu}


\begin{abstract}

A major outstanding question regarding the formation of planetary systems is whether wide-orbit giant planets form differently than close-in giant planets. We aim to establish constraints on two key parameters that are relevant for understanding the formation of wide-orbit planets: 1) the relative mass function and 2) the fraction of systems hosting multiple companions. In this study, we focus on systems with directly imaged substellar companions, and the detection limits on lower-mass bodies within these systems. First, we uniformly derive the mass probability distributions of known companions. We then combine the information contained within the detections and detection limits into a survival analysis statistical framework to estimate the underlying mass function of the parent distribution. Finally, we calculate the probability that each system may host multiple substellar companions. We find that 1) the companion mass distribution is rising steeply toward smaller masses, with a functional form of $N\propto M^{-1.3\pm0.3}$, and consequently, 2) many of these systems likely host additional undetected sub-stellar companions. Combined, these results strongly support the notion that wide-orbit giant planets are formed predominantly via core accretion, similar to the better studied close-in giant planets. Finally, given the steep rise in the relative mass function with decreasing mass, these results suggest that future deep observations should unveil a greater number of directly imaged planets.

\end{abstract}

\newpage
 
\section{Introduction}


Recent high-contrast imaging surveys of nearby stars have began to unveil a population of wide-orbit ($a\geq8$ AU) giant companions that are unlike anything found in our solar system. These companions are typically at least twice the mass of Jupiter and twice its orbital separation. Some objects are within bounds of being planetary companions (e.g., \citealt{Marois2008}, \citealt{Lagrange2010}, \citealt{Rameau2013}, \citealt{Macintosh2015}, \citealt{Chauvin2017}, \citealt{Keppler2018}), while other, yet more massive objects, are among the class of brown dwarfs and low-mass stars (e.g., \citealt{Metchev2006}, \citealt{Kuzuhara2013}, \citealt{Konopacky2016}, \citealt{Milli2017}, and others). As an ensemble, these wide-orbit companions enable us to study the formation of outer planetary systems in a way that is similar and complementary to the prevalent studies of inner planetary systems.\footnote{For a recent review of directly imaged planetary mass companions, see \cite{Bowler2016}.}

Two main mechanisms for the formation of wide-orbit giant companions within protoplanetary disks have been suggested and explored: 1) top-down formation by gravitational disk instability (GI: e.g., \citealt{Boss1997}), and 2) bottom-up formation by core accretion (CA: e.g., \citealt{Pollack1996}). While both mechanisms may plausibly operate within protoplanetary disks, they are expected to produce very different signatures in the statistics of companion properties (for example, see the population synthesis studies of \citealt{Mordasini2009}, \citealt{Forgan2018}, \citealt{Mul2018}, and others). For companions not born in disks, collapse within the protostellar core phase is a plausible option, and the distribution of companion masses would likely resemble the low-mass end of the stellar initial mass function (e.g., \citealt{Kroupa2001}, \citealt{Chabrier2003}).

The criteria of being less than the $\sim$13 M$_{\rm Jup}$ deuterium burning limit is a commonly used dividing line between planets and brown dwarf companions. This is often scrutinized, in part because it is not a formation-motivated definition. In this study we will treat both classes of objects uniformly. We will refer to objects from both categories as ``wide-orbit companions", while for simplicity we will frequently refer to those beneath the deuterium burning limit as ``planets", and those above this limit as ``brown dwarfs". We make no further distinction in our definitions based on orbital configuration. 

With these definitions in mind, we now turn to briefly summarize the physical processes of GI and CA, focusing on their expected contributions to the relative frequency of planets and brown dwarf companions, and expected fractions of systems with multiple companions.

Theoretical simulations suggest that GI typically produces very massive companions, and operates more easily at larger separations (e.g., \citealt{Matzner2005}, \citealt{Rafikov2005}, \citealt{Clarke2009}, \citealt{Kratter2010}). As a result, the majority of GI-born companions are likely massive enough to be classified as brown dwarfs and low-mass stars, with the process yielding a small (but perhaps detectable) fraction of planetary-mass companions. This is due to the fact that the process must begin very early, while enough mass exists in the disks to trigger the instability. In turn, this causes the majority of objects formed by GI to grow rapidly in mass, given the availability of material at such young ages (e.g., \citealt{Kratter2010}, \citealt{Forgan2013}, \citealt{Forgan2018}). Still, it is possible that some planetary mass companions may originate from disk-born GI, which would be evident in a lower (or consistent) frequency of giant planets compared to brown dwarfs. While it is plausible that GI could produce multiple companions in the same system, overall the mechanism is expected to yield a low multiplicity fraction \citep{Forgan2018}. 

On the other hand, formation of giant planets via CA involves much longer timescales, primarily limited by the time required for the solid core to grow above the critical mass to trigger runaway gas accretion \citep{Bodenheimer1986}. Typically, $\sim$10 M$_{Earth}$ is considered as the critical core mass, although recent work has shown that smaller masses (down to several M$_{Earth}$) are sufficient to trigger runaway gas accretion at larger disk radii (\citealt{Piso2014}). Additional factors, such as pebble accretion (e.g., \citealt{Ormel2010}, \citealt{Lambrechts2012}), may help to accelerate solid core growth. Nevertheless, the growth of a massive core and subsequent accumulation of a gaseous envelope is in contest with the dispersal of the gaseous protoplanetary disk ($\lesssim$10 Myr; \citealt{Ercolano2017}). The late formation of the cores within a disk rapidly declining in mass limits the availability of accretable gas and, thus, the probability of the formation of super-Jupiters and brown dwarfs. As a result, and contrary to the GI scenario, CA is expected to produce a much higher relative frequency of lower mass planets compared to super-Jupiters and brown dwarfs. 

Furthermore, the fraction of systems hosting multiple giant companions is much higher for close-in planets formed via CA. \cite{Knutson2014} studied 51 systems containing giant planets of 1-13 M$_{\rm Jup}$ at orbital separations between 1-20 AU and found that the occurrence rate of additional massive outer companions is 51$\pm$10\%. Similarly, the fraction of planetary systems hosting confirmed multiple planets is $\geq$21.8\% and $\geq$24.3\%, for detection via transit and radial velocity, respectively. The lower limit is the confirmed fraction of multiple systems, and the true fraction of multiple systems among these is likely even higher, considering that additional planets may exist that are either non-transiting or of sufficiently long-period to be non-detected.\footnote{Data obtained from exoplanet.eu \citep{Schneider2011} on January 4, 2019.} If the wide-orbit giant planets also formed via CA, we might expect a significant fraction of these systems to host multiple giant companions. 


The prevalence of binary and multi-star systems is evidence that protostellar cores are frequently subject to fragmentation. However, as the evolving system continues to decline in mass, the probability that GI will occur in the disk stage is diminished. Likewise, the resulting companion mass is also limited by the availability of material at later times, with the most likely outcome of disk-born GI being a companion in the brown dwarf regime. The paucity of such companions to main-sequence stars confirms this general picture. This trend, known as the ``brown dwarf desert", was initially identified among close-in companions (e.g., \citealt{Duq1991}). The same trend has also been identified in the low occurrence rates of wide-orbit brown dwarf companions (e.g., \citealt{McCarthy2004}, \citealt{Kraus2011}, \citealt{Vigan2017}), although the effect is not as extreme, with a few percent of stars hosting such companions \citep{Metchev2009}.

Similarly, relatively few directly imaged giant planets have been discovered. For this reason, there is an on-going debate over the dominant formation pathway of these objects, with arguments in favor of both GI$-$ and CA-like processes. While significant difficulty remains in determining the formation pathway of a particular object, the dominant formation mechanism for an ensemble of objects can be revealed by the form of their relative mass function. If the mass distribution reveals a higher relative frequency of lower-mass objects, this would indicate that similar CA-like planet formation processes occur within the inner and outer regions. On the other hand, if the mass function is relatively flat, or rising toward higher masses, this would indicate GI as the dominant formation mechanism. Likewise, insight may be gained by examining the possibility that a significant fraction of systems may host multiple giant companions, as CA is expected to produce a significant fraction of such systems. 




Here, we aim to constrain the relative mass function and multiplicity of directly imaged wide-orbit giant companions by applying the class of statistical methods that were developed for analyzing censored data. These methods, often referred to as ``survival-analysis", are well-vetted in medical and risk management industries, and are becoming increasingly popular in astronomy. In the case of directly imaged companions, the data comprise a set of detected objects with estimated masses\footnote{The masses are typically estimated via the combination of the system's age and distance, the companion's photometry, and a model grid that describes the mass-luminosity-age evolution.}, and a population of lower mass objects that are possibly present, but non-detected, that are included as upper mass limits.  In particular, we will utilize the Kaplan-Meier maximum likelihood estimator (\citealt{Kaplan1958}, \citealt{Feigelson1985}) to estimate the cumulative distribution of the underlying parent population. 

In this study, we focus on systems with known wide-orbit substellar companions. With this approach, we isolate the question of what is the \textit{relative} mass function of wide-orbit companions in systems where they have been identified, from the question of whether these systems are exceptional (i.e., having an overall low occurrence rate). This approach is different from, but complementary to the conventional occurrence rate studies, in which a mass function is typically assumed and then contrasted with the observed detection rate. In the conventional approach, the mass function and the occurrence rates are degenerate, and both are unknown (e.g., \citealt{Kasper2007}, \citealt{Biller2013}, \citealt{Brandt2014}, \citealt{Nielsen2013}, \citealt{Galicher2016}, \citealt{Reggiani2016}, \citealt{Vigan2017}, \citealt{Stone2018}, \citealt{Nielsen2019}). 

We begin by assembling a list of known companions in \S2. We then derive their mass and upper mass limit probability distributions in \S2.1. We describe how these mass measurements and mass limits are incorporated into the survival analysis framework in \S2.2. In \S3, we present our results, namely the relative mass function of giant companions in \S3.1, and the associated multiplicity probabilities in \S3.2. We explore the effects of various model assumptions in \S3.3 \& \S3.4, and show results for a selection of relevant subsamples in \S3.5. Finally, we provide a brief discussion of the results and a critical assessment of the weaknesses of our approach in \S4, and conclude by summarizing our findings in \S5.



\section{Sample of Companions}

We assembled a list companions from the literature, beginning with the list on \texttt{exoplanet.eu} \citep{Schneider2011} as of 2018 November 23 with the selection criteria of being discovered by direct imaging. While this list is often scrutinized, in particular for containing objects beyond the deuterium burning limit as ``planets", for our purposes this is desirable, as we aim to constrain the mass distribution across the planet to brown dwarf mass threshold. To our knowledge, this list is complete with respect to companions that have been directly detected in high-contrast imaging with mass estimates in the sub-stellar range, which we verified by cross-checking against the NASA Exoplanet Archive.\footnote{http://exoplanetarchive.ipac.caltech.edu} We excluded planetary-mass companions around white dwarf and brown dwarf hosts, and also planet candidates that have been interpreted as potential disk features (e.g., the planet candidates around Fomalhaut: \citealt{Janson2012}, \citealt{Lawler2015}, HD 169142: \cite{Ligi2018}, and LkCa 15: \citealt{Thalmann2015}, \citealt{Mendigutia2018}). This criteria resulted in a list of 57 companions, whose properties are given in Appendix A.

The host stars among this sample, and the properties of their companions, are highly diverse. The host stars range from ages of a few Myr to several Gyr, and display spectral types spanning late-M to early-A types ($\sim$0.2$-$3 M$_{\odot}$ for main sequence stars), which reflects the diversity of selection criteria among the original surveys. Furthermore, their companions have estimated masses ranging from $\sim$2 M$_{\rm Jup}$ to the minimum hydrogen burning mass, and occupy orbital ranges of 8 AU to several thousand AU (estimated from their projected separations, in most cases). To reduce potential effects of including such a variety of orbital properties and host star mass, we restricted our initial analysis to companions whose projected separation is $\leq$100 AU, and with host stars of spectral type A0-K8 such that the mass ratio is $\lesssim$0.01 for a 5 M$_{Jup}$ planet around the least massive stars. This resulted in a list of 23 companions. We refer to this population as ``sub-sample A", or our primary sample, and will discuss results obtained from this population unless otherwise noted. In \S3.5 we will relax these criteria and examine the full sample, and will also examine select subsamples to utilize the full diversity among our sample to search for trends in companion properties.


\subsection{Conversion of Photometry to Masses and Upper Mass Limits}

For each companion, we compute its mass probability distribution via a Monte Carlo (MC) simulation drawing from Gaussian priors on age, distance, and photometry, as reported in the literature. We convert these properties into mass estimates via the evolutionary tracks of \cite{Baraffe2003} for our primary analysis, and explore other models (including dusty photospheres, and ``cold-start" initial conditions) in \S3.4. In general, we utilize the most sensitive measurements currently available for upper limits of additional companions. We consider detection limits only in the outer regions, which are well-matched to the wide-orbit population that is the focus of this study. In these regions (typically $\gtrsim$0$\farcs$5), the sensitivity is not primarily limited by speckle noise from the central star, and instead is limited by thermal background and other spatially homogeneous sources of noise. When sensitivity between photometric bands is comparable, we utilize the longest wavelength data available (typically either $Ks$ or $L^{\prime}$) since these are less likely to be affected by differences in molecular absorption. 

For systems without published detection limits, we estimated 5$\sigma$ detection limits by assuming that the photometric uncertainty on the known companions corresponds to the $\sim$1$\sigma$ noise level.\footnote{The formula used for this conversion, and the systems for which it has been applied, can be found in Table A1.} We tested this method of estimating detection limits on companions with published limits and found that this method consistently over-estimates the upper limits due to the fact that the photometric uncertainty also incorporates photon noise from the detection (which can be quite high), whereas a true detection limit would be dominated by background noise terms. In other words, these are likely conservative estimates on the detection limits within these systems. We use the most up-to-date age ranges available throughout the literature, and assume a Gaussian profile within this range.\footnote{Where necessary, we convert asymmetric uncertainties into a symmetric age range.} Likewise, we utilize the most up-to-date distance measurements available, which typically come from \textit{Gaia DR2} \citep{GDR2}. 

Given the sparse and non-uniform sampling of the evolutionary model grids, we must interpolate between the points in order to generate solutions at arbitrary masses and ages (though still within the bounds of the grids, which for the \citealt{Baraffe2003} models is 0.5-100 M$_{\rm Jup}$ and 1 Myr to 10 Gyr). We adopted a bi-linear interpolation scheme, and also tested the output of cubic interpolation. We found similar results in both cases (mass probability distributions of similar mean and width), and choose to retain the simpler bi-linear interpolation for the proceeding analysis.

We show the cumulative mass function of the detected objects around A0-K8 stars and with projected separations $\leq$100 AU in Fig 1. The individual mass probability distributions for the detected companions, and the mass detection limit probability distributions are shown in Appendix B, along with objects not included in the primary sample. Overall, the cumulative distribution shows a steeper slope toward lower masses, although an important (and non-physical) feature of this cumulative mass function is that it drops to zero below $\sim$2-3 M$_{\rm Jup}$, which simply reflects an observational bias arising from the difficulty of detecting such low-mass companions. In the next subsection, we estimate the correction to this distribution at small masses.

\begin{figure}[h]
\figurenum{1}
\epsscale{1.1}
\plotone{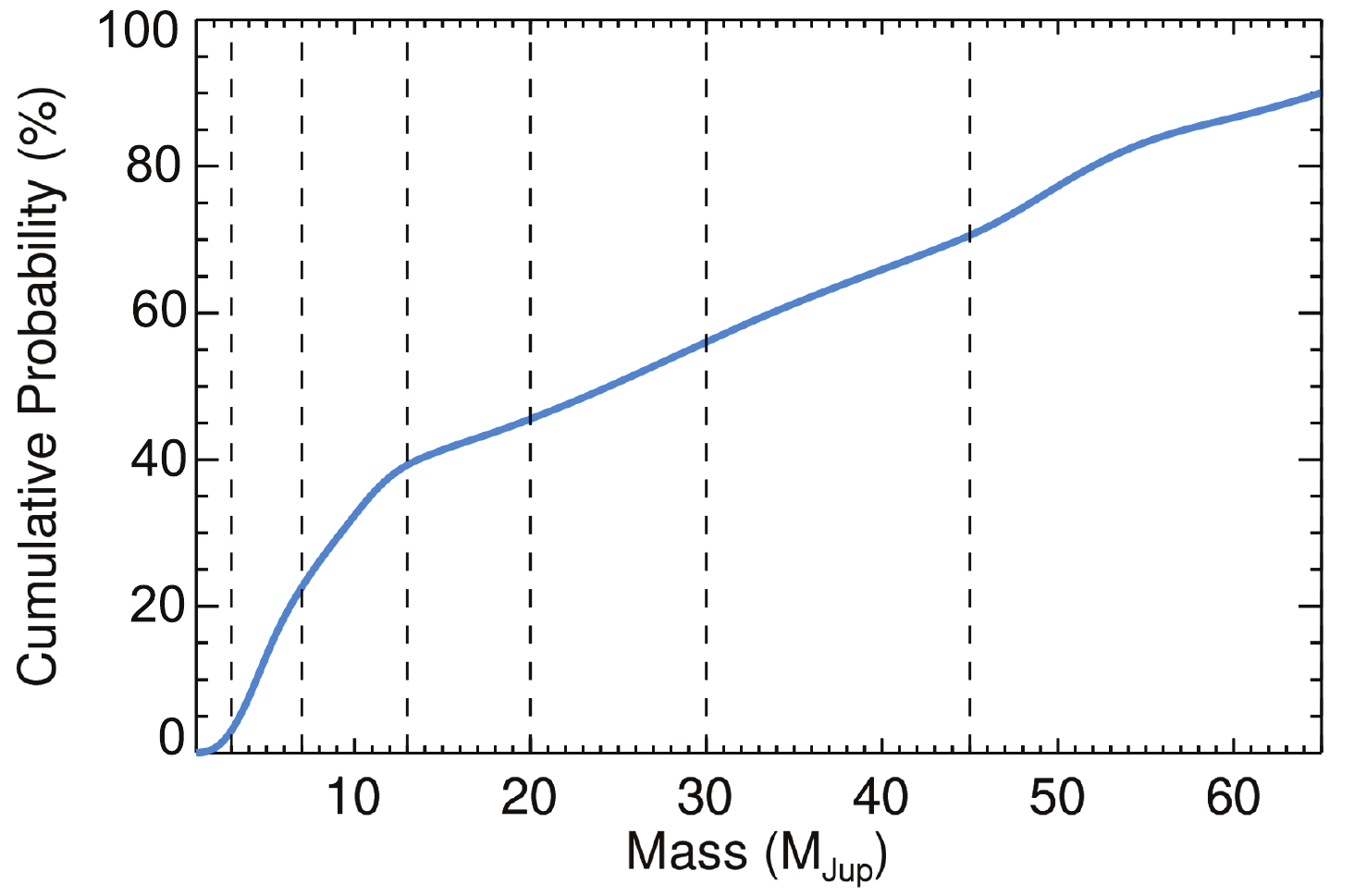}
\caption{The cumulative mass probability distribution of directly imaged companions within 100 AU of A0-K8 stars. Overall, the slope is steeper at lower masses, reflecting a higher relative frequency of objects detected with low mass compared to objects of higher mass. The dashed lines show the pre-selected mass bins that will be utilized in the proceeding analysis.}
\end{figure}


\subsection{Survival Analysis: Estimating the Underlying Cumulative Mass Probability Distribution}

The class of statistical methods that has been developed for dealing with censored data$-$e.g., data containing both detections and detection limits$-$is frequently referred to as ``survival analysis". While many works are devoted to exploring these methods in detail, we refer the interested reader to \cite{Feigelson1985}, which recasts the typical formulation from a context of right-censored data (involving lower limits, or in the name-sake problem, survival times) to a context of left-censored data (involving upper limits), which is applicable for our data-set, and in general for most astrophysical contexts.

We utilize the Kaplan-Meier (KM) maximum likelihood estimator (\citealt{Kaplan1958}, \citealt{Feigelson1985}), which approximates the cumulative distribution of the underlying parent population from which the censored data were drawn. Specifically, we utilize the form given in Section 2, Equations 1-8 of \cite{Feigelson1985} for a sample containing indistinct measures (in this case, multiple objects within the same mass bin). The general form of the KM estimator is a monotonic increasing function whose value only changes at the values of uncensored measurements, with the size of the jumps being determined by the combination of the censored and uncensored measures. In this way, the KM estimator provides an estimated correction for the observational bias at low masses by including the information contained within the detection limits. While the result is likely closer to reality than considering merely the detections alone, it remains an approximation since the true masses of the undetected companions, and the number of such companions that actually exist beneath the detection limits, remain unknown (this is a topic of further discussion in \S3.1 and \S4.1). 

\begin{figure}[h]
\figurenum{2}
\epsscale{1.1}
\plotone{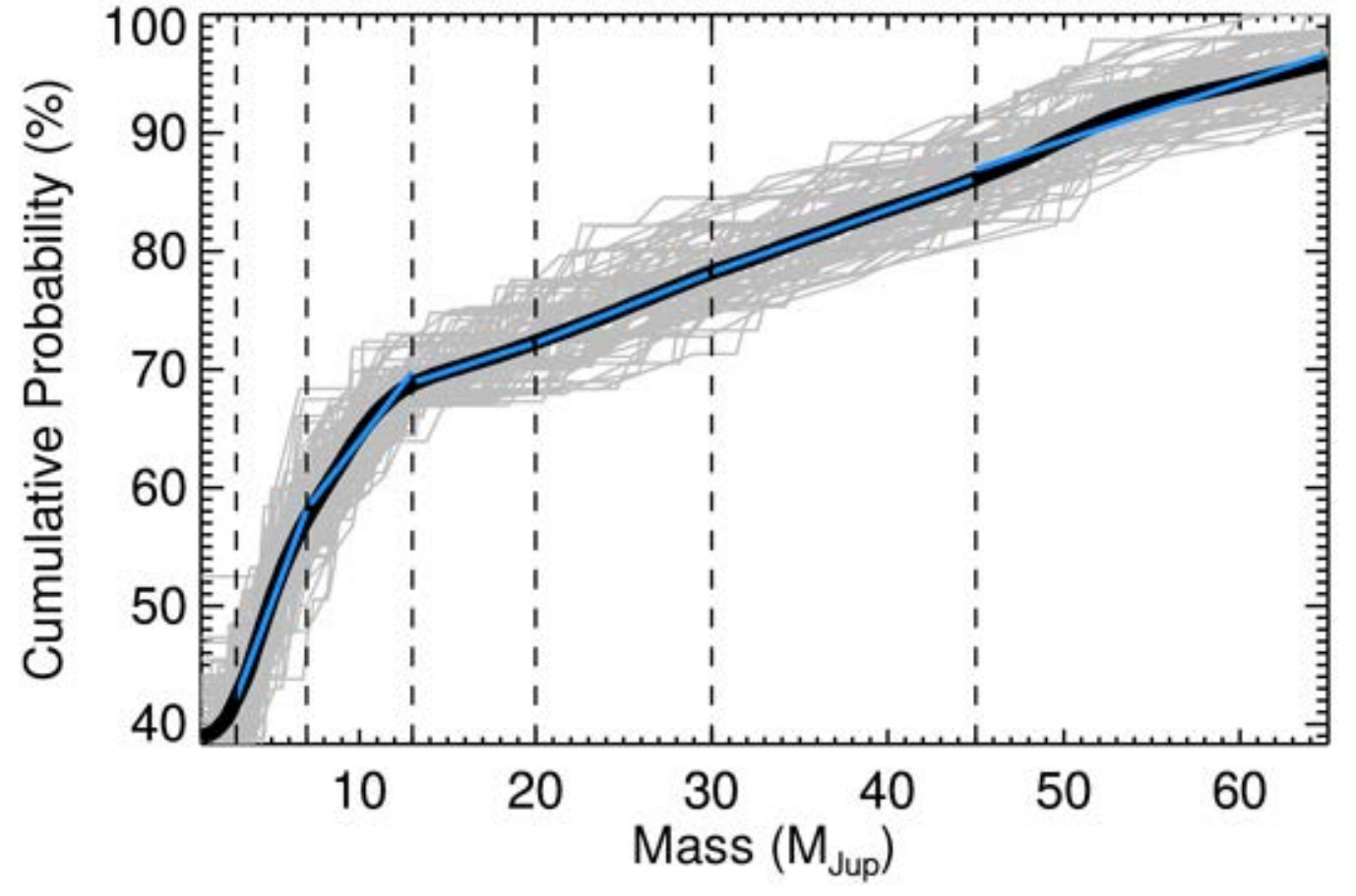}
\caption{The cumulative mass probability distribution of directly imaged companions within 100 AU of A0-K8 stars resulting from the survival analysis. Here we show the distributions of 100 randomly selected MC trials (gray) and the average of 1,000 trials (black). The dashed lines show the pre-selected mass bins. In each bin, we determine the best linear fit (blue). The ratio of slopes between fits to different bins provides the relative frequency.}
\end{figure}

While the measurements originate from (often very) different data-sets, the end products are the same: namely photometry of detected sources and photometric detection limits on additional sources. By uniformly converting these measurements into estimated masses and mass detection limits, we eliminate the possibility that differing methods of converting the original measurements into estimated masses may bias our results. There exists the possibility that differing strategies for estimating photometric sensitivity may lead to different results (e.g., \citealt{Mawet2014}). However, these effects are most prevalent at small separations, whereas we consider detection limits only in the outer regions, in which the sensitivity does not vary significantly with angular separation, and in which the noise is approximately Gaussian. Any remaining differences in the original data reductions are likely to enter as random errors, and on average should not bias our results.

To incorporate the measurement uncertainties, we compute the survival function via an MC simulation of 1,000 trials, wherein each trial we calculate the KM estimator by randomly drawing a mass and upper mass limit from each companion's probability distributions.\footnote{In this way, we implicitly assume that each detection limit corresponds to one non-detected object. On average, this is likely a reasonable approximation, and we will discuss the effect of altering this assumption in \S3.1 and \S4.} We then average the cumulative distributions together, which is shown in Fig. 2 along with the cumulative distributions of 100 randomly selected MC trials. We split the distribution into six mass bins (3-7, 7-13, 13-20, 20-30, 30-45, and 45-65 M$_{\rm Jup}$), and within each bin fit a linear model.\footnote{These mass bins are selected to roughly coincide with the inflection points in the distribution. The exact selection of mass bins does not significantly affect the results.} We repeat this analysis on each of the 1,000 MC trials to establish uncertainties on the relative frequency of each mass bin. The relative slopes of the linear fits provides an estimate of the relative frequency of companions within these mass bins, which is the topic of the following section.


\section{Results}

\subsection{The Wide-Orbit Planetary Mass Function}

The derivative of the cumulative mass function provides the relative mass function, which we henceforth refer to as the companion mass function (CMF). In Fig. 3, we show the result derived from the survival analysis-generated cumulative mass function (blue points), alongside the result derived from the cumulative probability distribution of the detections alone (red points). In both cases, the distribution drops steeply between the first three mass bins (3-20 M$_{\rm Jup}$), and is relatively flat at higher masses, with an approximate functional form of $N \propto M^{-1.3\pm 0.3}$. The similarity between the two distributions is due to the fact that many of the detection limits are lower than the minimum mass of 3 M$_{\rm Jup}$ considered here, which mostly shifts the cumulative distribution upward without affecting its overall shape. Nevertheless, some detection limits are higher than 3 $M_{\rm Jup}$, which can be seen as a higher relative frequency of 3-7 and 7-13 M$_{\rm Jup}$ objects in the distribution resulting from the survival analysis. 

The magnitude of this difference is impacted by our assumption that each detection limit corresponds to a \textit{single} object whose mass is beneath the detection limit. If \textit{multiple} companions exist within any of these systems that are beneath the detection limits, that would increase the frequency of planetary-mass companions even further. On the other extreme, if there is not a single companion beneath the detection limits among any of these systems, then the distribution would follow that of the detected objects alone, which sets a lower limit to the change of the slope across the CMF. Given this bottom-heavy CMF, we expect that more non-detected companions exist in the lower-mass range, and thus in the proceeding sections we utilize the results derived from the survival analysis.

\begin{figure}[t]
\figurenum{3}
\epsscale{1.02}
\plotone{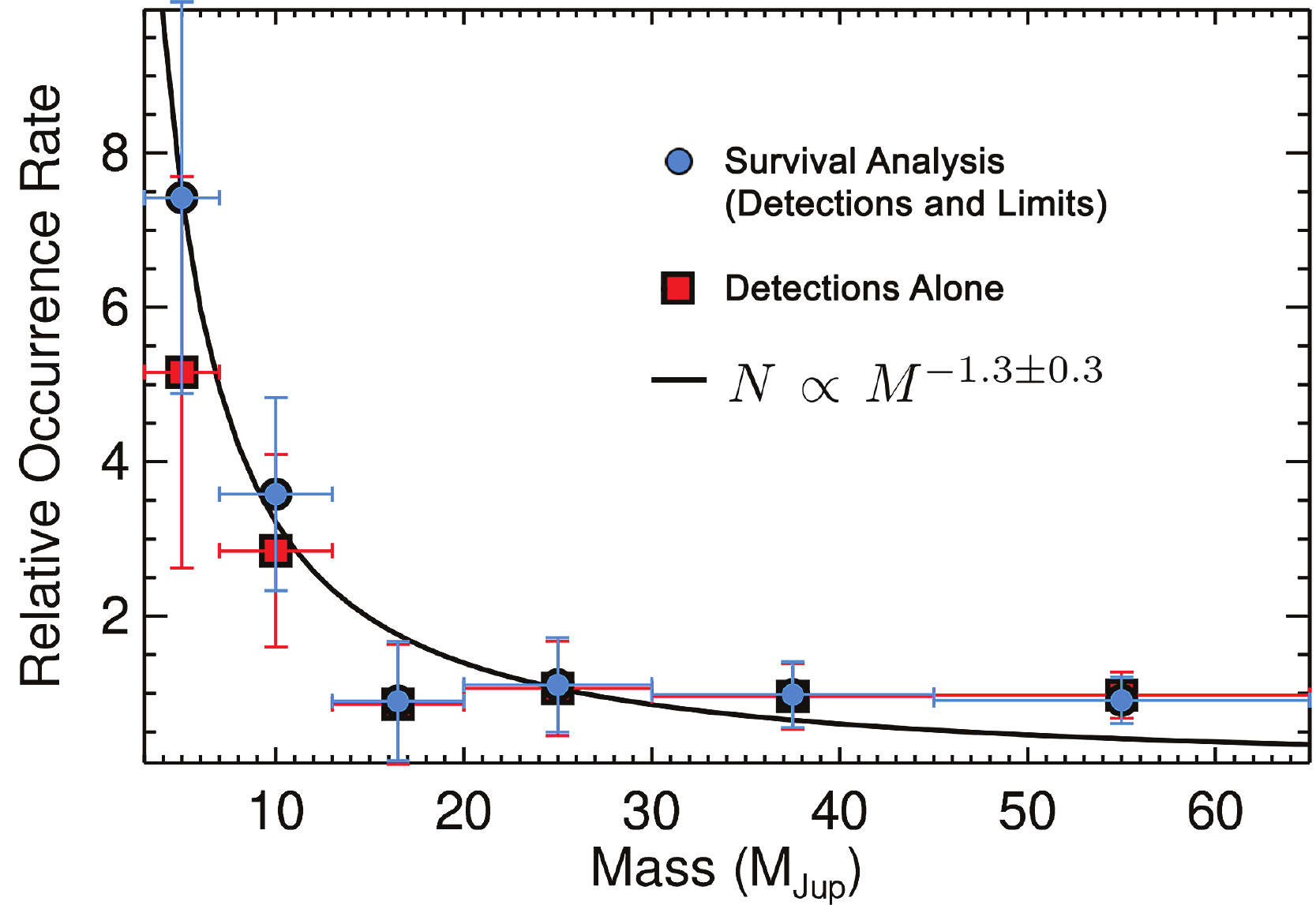}
\caption{The relative frequency of wide-orbit companions as a function of mass derived from survival analysis methods (blue points) and from the detections alone (red points). Both distributions are bottom heavy and correspond approximately to a power law of $M^{-1.3}$. The difference between the two distributions is a slight enhancement at low-masses in the survival analysis-derived distribution, which is representative of the typically very sensitive detection limits (i.e., most non-detected companions must be low-mass).}
\end{figure}


\subsection{Comparison to Other CMFs}


In Fig. 4 we compare the observed CMF of wide-orbit companions to simulated CMFs from theoretical models of companions produced through solely CA \citep{Mordasini2009} and by GI \citep{Forgan2018}. We also compare the results to the CMF of inner planets discovered by RV surveys \citep{Schneider2011}. The CA population synthesis of \cite{Mordasini2009}, and the relative frequency of close-in giant planets as derived from RV surveys\footnote{Data obtained from exoplanet.eu on November 7, 2018.} are in good agreement with the data (unreduced $\chi^2 \sim$ 5-6 in both cases). A similar population synthesis model with GI as the dominant formation mechanism \citep{Forgan2018} does not match the data, as it follows a relatively flat distribution ($\chi^2 \gtrsim$ 100). 

However, when the GI model is re-normalized to fit only the highest three mass bins (20-65 M$_{\rm Jup}$), we see that this model does a fair job at matching the high-mass end of the distribution ($\chi^2 \sim$ 1 with respect to only these points), while contributing less significantly to the relative abundance of planetary mass companions (roughly 6\% of the 3-7 M$_{\rm Jup}$ bin, and 14\% of the 7-13 M$_{\rm Jup}$ bin). Similarly, we compared the results to a low-mass stellar initial mass function (IMF; \citealt{Kroupa2001}). The result is essentially the same as for the GI model, which is to be expected since both distributions are relatively flat. While the stellar IMF is a poor fit to the distribution throughout the complete range of masses ($\chi^2 \sim$ 20), it provides an equivalent match to the higher-mass objects as the GI model. 

\begin{figure}[t]
\figurenum{4}
\epsscale{1.1}
\plotone{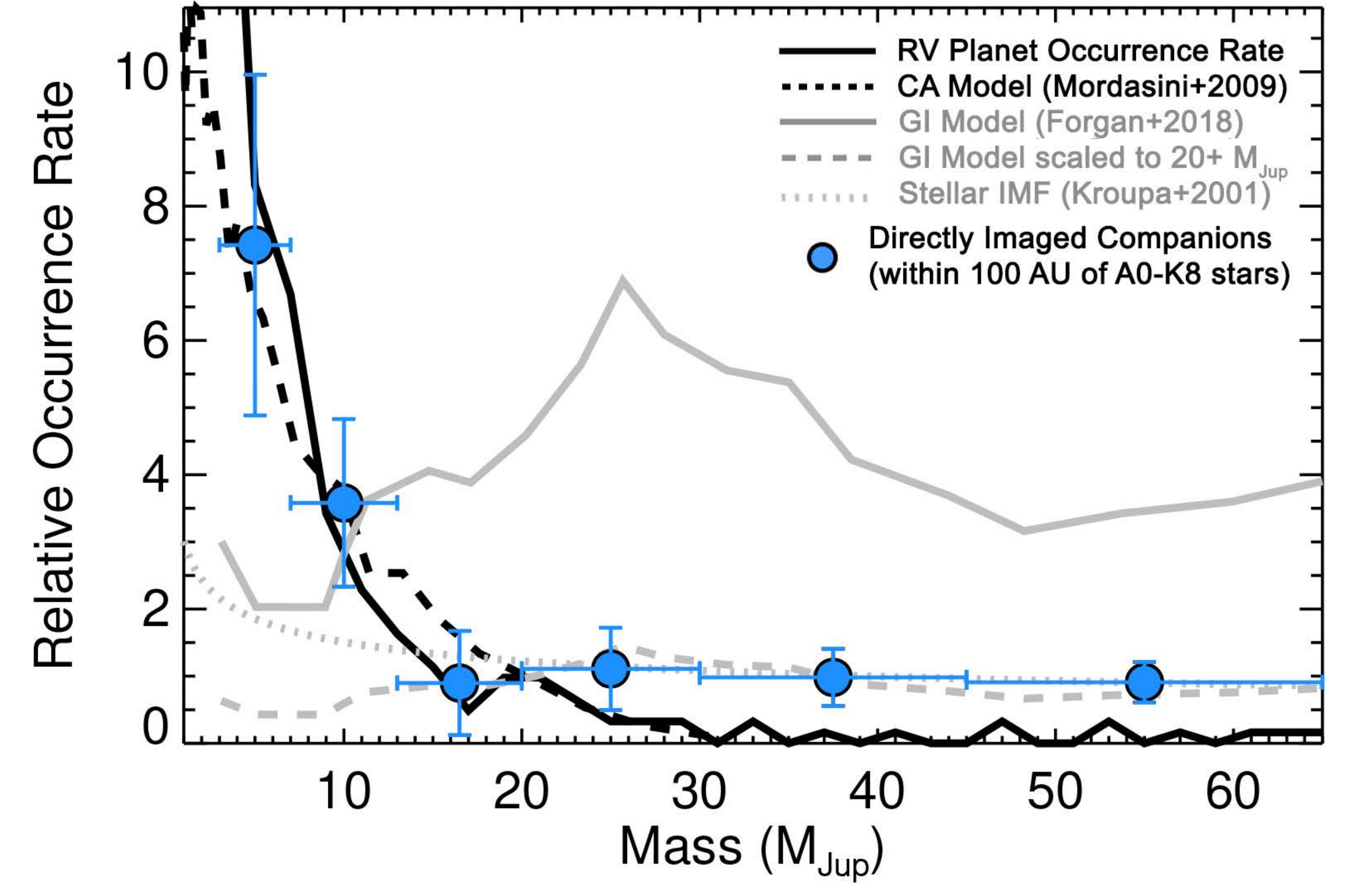}
\caption{The wide-orbit companion mass function (CMF; blue points). Simulated CMFs resulting from population synthesis models for core accretion (CA) and gravitational instability (GI) followed by tidal downsizing are shown in black and gray, respectively. The observed CMF of planets detected by radial velocity (RV) is shown in the blue curve. An extension of the Kroupa IMF is shown in the dotted gray curve. Each distribution is normalized to the wide-orbit CMF by the mean of the individually calculated normalization factors for each mass bin, weighted by the inverse of the measurement uncertainties. The dashed gray curve shows the GI population synthesis model normalized to only the objects greater than 20 M$_{\rm Jup}$.}

\end{figure}

\subsection{Multiplicity Probabilities}

The second aim of our study is to assess the fraction of these systems that could host multiple wide-orbit giant companions. So far HR 8799 and HIP79930 are the only examples systems with multiple companions that have been detected, although other candidate multiple systems exist. HR 8799 is a remarkable system containing \textit{four} super-Jupiters between 10-70 AU (\citealt{Marois2008}, \citealt{Marois2010}). HIP73990 is also a remarkable case, as it hosts two brown dwarfs at projected separations of $\sim$18 and 28 AU \citep{Hinkley2015}. Since the actual number of observed multiples is low, the question of multiplicity essentially becomes whether the (apparently single) systems are in fact compatible with hosting additional companions that are beneath the detection limits. 

To address this possibility, we begin with the assumption that each system hosts an additional companion whose mass is independent of other bodies in the system. In reality, systems hosting one wide-orbit giant companion may be more (or less) likely to host additional companions. In the following, we explore the simplest scenario in which the masses are independent. If future, deep searches fail to reveal more companions within the known systems, then this could be taken as evidence that giant companion formation inhibits the potential for forming a second companion. On the other hand, if more systems are discovered to be multiples than predicted here, this would suggest that systems that are able to form a single giant companion have a higher likelihood of forming multiple such companions.

We also assume that the hypothetical second companion exists within the semi-major axis range corresponding to that in which the upper mass limit was defined, which for simplicity can always be taken to be external to the known companion. We then perform an MC simulation of 1,000 trials, wherein each trial we randomly draw a second companion mass from the CMF, and an upper mass detection limit from the probability distributions derived in \S2.1. We approximate an upper limit to the probability that each system hosts an additional wide-orbit super-Jovian companion by the fraction of trials resulting in a randomly drawn companion whose mass is beneath the detection limit for that system and $\geq$2 M$_{\rm Jup}$.

\begin{deluxetable}{cccc}
\tabletypesize{\scriptsize}
\tablecaption{Multiplicity Probabilities}
\tablewidth{0pt}
\tablehead{\colhead{System} & \colhead{P(Double) \%}  & \colhead{P(Triple) \%}  & \colhead{P(Quad) \%} }
\startdata
51 Eri & 34.1 & 11.6 & 3.97 \\ 
GJ 504 & 77.9 & 60.7 & 47.3 \\ 
GJ 758 & 85.6 & 73.3 & 62.7 \\ 
HD 1160 & 79.1 & 62.6 & 49.5 \\ 
HD 19467 & 94.1 & 88.5 & 83.3 \\ 
HD 206893 & 84.2 & 70.9 & 59.7 \\ 
HD 4113 & 89.3 & 79.7 & 71.2 \\ 
HD 95086 & 40.9 & 16.7 & 6.84 \\ 
HD 984 & 76.2 & 58.1 & 44.2 \\ 
HIP 65426 & 53.3 & 28.4 & 15.1 \\ 
HIP 73990 & 100. & 79.6 & 63.4 \\ 
HIP 74865 & 74.3 & 55.2 & 41.0 \\ 
HR 2562 & 68.3 & 46.6 & 31.9 \\ 
HR 3549 & 60.8 & 37.0 & 22.5 \\ 
HR 8799 & 100. & 100. & 100. \\ 
PDS 70 & 65.7 & 43.2 & 28.4 \\ 
PZ Tel & 48.4 & 23.4 & 11.3 \\ 
$\beta$ Pic & 22.5 & 5.06 & 1.14 \\ 
$\kappa$ And & 40.8 & 16.6 & 6.79 \\ 
\hline\\

Mean & 68.2\% & 50.4\% & 39.5\% \\

\enddata

\tablecomments{Note: this table also serves to identify the companions that were considered as part of our primary analysis (those within 100 AU of A0-K8 stars), and is a subset of the objects whose properties are described Table 3.}
\end{deluxetable}


These probabilities are given in Table 1, along with the probabilities that each system may host two and three additional companions via the same reasoning. On average, these systems have a 68.2\% probability of hosting a second wide-orbit giant companion ($\geq$2 M$_{\rm Jup}$) drawn from the CMF. Likewise, on average there is a 50.4\% probability of hosting three such planets, and a 39.5\% probability of hosting four planets, like the HR 8799 system. The frequency of similar systems with multiple super-Jupiter companions will be a topic of discussion in \S4.6.

These results are likely overly optimistic about the fraction of multiple systems, and could instead be considered as upper limits to the probability of hosting additional wide-orbit giant companions. In particular, semi-major axis effects also likely play a significant role in the probability that additional companions may exist. For instance, by considering requirements for dynamical stability of multiple orbiting bodies, these probabilities may be reduced further. While it is possible to increase the complexity of this analysis to include such effects, this simple analysis is revealing enough for our purposes: the observed bottom-heavy CMF in combination with the available detection limits suggests that some of these systems are likely hosting wide-orbit planetary mass companions that have not yet been detected.


\subsection{Exploration of Model Assumptions}

The above results were derived under the assumptions inherent in the \cite{Baraffe2003} models: namely that planets and brown dwarfs retain all of their initial entropy (representative of a ``hot-start" formation scenario) and have clear atmospheres. Now, we explore the effect of relaxing those assumptions. In \S3.4.1 we explore the effect of planets forming with a variety of initial entropy conditions (representative of a ``cold-start" scenario). In \S3.4.2, we explore the effect of allowing companions to retain a significant fraction of dust in their atmospheres by utilizing the grid of models presented in \cite{Chabrier2000}.

\subsubsection{Hot vs. Cold-start Planets}

The initial luminosity of young giant planets remains an open question of giant planet formation (e.g., \citealt{Marley2007}, \citealt{Fortney2008}, \citealt{Spiegel2012}, \citealt{Mordasini2017}). The uncertainty primarily lies in how much energy is radiated away from the in-falling material at the accretion shock boundary \citep{Marleau2017}. The radiative efficiency is not clearly predicted from simulations, and while some young companions display an observable accretion luminosity (e.g., \citealt{Zhou2014}, \citealt{Close2014}, \citealt{Sallum2015}, \citealt{Wagner2018}), significant difficulty remains in establishing a radiative efficiency from these limited observations. 

The limiting conditions in which the shocked gas radiates 0\% and 100\% of its kinetic energy at the shock boundary lead to the classical ``hot" and ``cold" start models (e.g., \citealt{Baraffe2003}, \citealt{Marley2007}). The reality is likely somewhere in between, and there probably exists a spread in initial luminosities for a given mass (\citealt{Mordasini2017}, \citealt{Berardo2017}). At higher masses, the hot and cold-start models converge as deuterium burning begins to contribute to the overall luminosity. At later ages, the hot and cold-start models also converge, with lower-mass tracks converging within a few tens of millions of years, and the $\sim$10 M$_{\rm Jup}$ tracks taking the longest to converge at a few hundred million years \citep{Spiegel2012}.

Our choice of utilizing the hot-start models of \cite{Baraffe2003} in the preceding analysis is motivated by the fact that for most of the mass range considered here, the hot and cold-start models are in good agreement (\citealt{Spiegel2012}, \citealt{Mordasini2017}). Our choice was further motivated by several indications suggesting that planets should form with initial luminosities close to those of the the hot-start models. For example, the planets in HR 8799 have dynamical masses that agree very well with the hot-start mass estimates (\citealt{Snellen2018}, \citealt{Wang2018}), and for $\beta$ Pictoris the presence of the disk is inconsistent with cold-start estimates for the planet's mass (\citealt{Lagrange2010}).  Furthermore, recent simulations \citep{Mordasini2017} have shown that, while the accretion shock may radiate away a significant amount of energy, continued accretion of planetesimals during this phase will lead to a luminosity-core-mass effect, whereby higher mass cores lead to higher initial luminosities. This effect causes the initial luminosities of cold-start planets to become comparable to those of the hot-start models with no core-mass effect. Nevertheless, this assumption may have an important effect on our final results, and deserves exploration. 

While cold-start evolutionary grids exist (e.g., the models of \citealt{Spiegel2012}), they remain significantly unconstrained due to the uncertainty in initial conditions and subsequent accretion history. Instead, we choose to employ a simple prescription to scale the luminosity of the hot-start evolutionary grids as an approximation of a cold-start case for objects beneath the deuterium burning limit. In computing the companion mass and upper mass limit probability distributions of objects, we assume a minimum efficiency representative of energy transfer during the accretion of the gaseous envelope. This enters as a numerical scaling factor in the temperature of the object, which we use to scale the luminosity by the corresponding fourth power of the change in temperature from the hot-start evolutionary grids. We assume a uniform distribution between this minimum efficiency and unity in the MC trials to assemble the mass probability distributions. This is representative of planets that formed in a variety of conditions, and is consistent with some fraction of planets attaining initial luminosities that are close to the hot-start predictions.

\begin{figure}[t]
\figurenum{5}
\epsscale{1.1}
\plotone{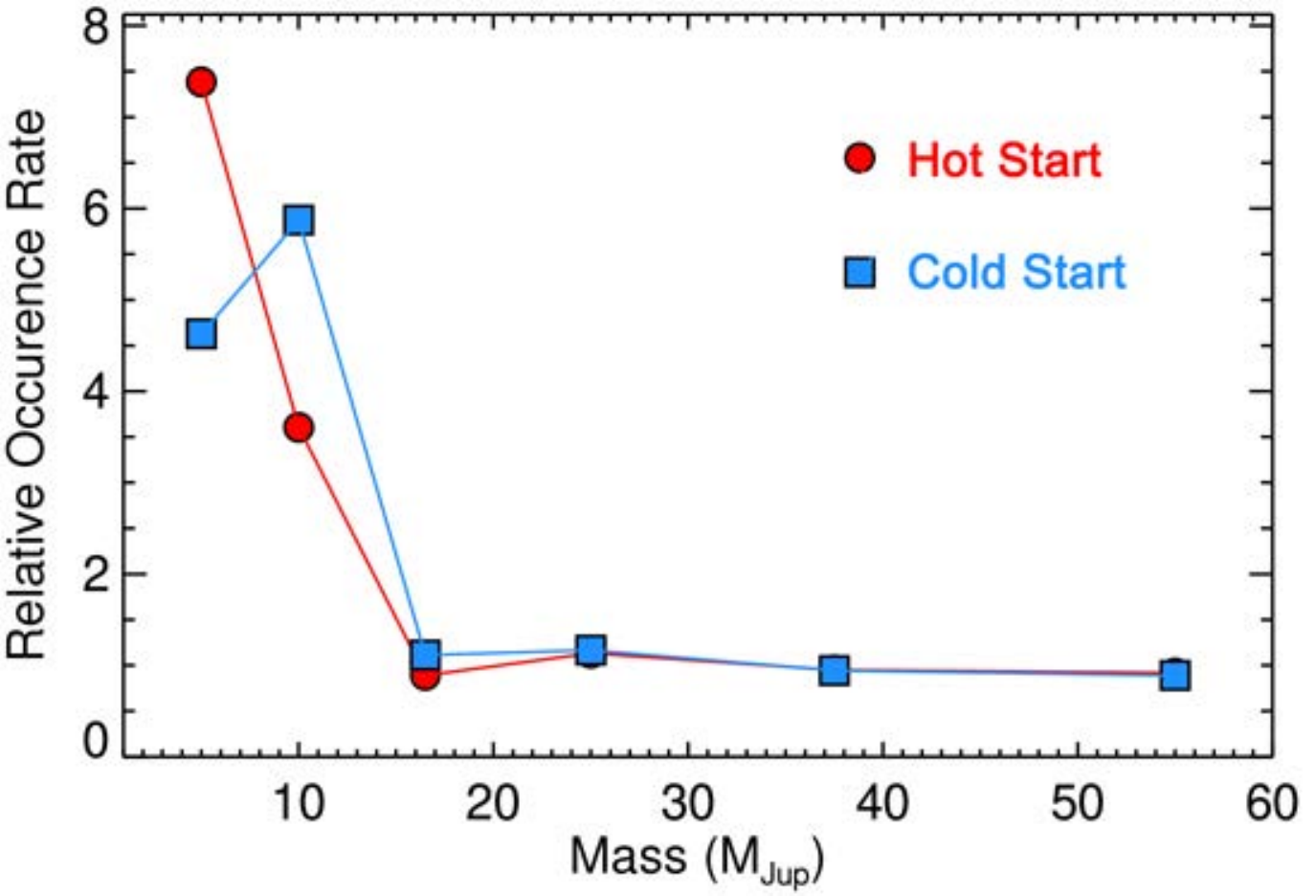}
\caption{The CMF for hot and cold-start assumptions of initial planetary luminosity. In both cases, the mass function is bottom heavy, with the steepest mass function representing the case in which planets retain most of their initial entropy (the hot-start case). The uncertainties are similar to those in Fig. 3 \& Fig. 4.}
\end{figure}

In Fig. 5, we show an example of the CMFs obtained for a minimum efficiency of 50\% corresponding to minimum luminosities that are 6\% of the maximum for a given mass. Assuming that planets form colder tends to move planets from the 3-7 M$_{Jup}$ bin to the 7-13 M$_{Jup}$ bin, but the latter are not moved beyond the deuterium burning limit due to convergence with the hot-start tracks at higher masses (\citealt{Spiegel2012}, \citealt{Mordasini2017}). In other words, under any assumption of initial planet entropy the result is still a bottom-heavy mass function. The general trend remains the same, with planetary mass companions being much more frequent than brown dwarfs. Similarly, the average upper limits on the probability that each system may host multiple companions drawn from the CMF are only slightly lower, with $\lesssim$48\%, $\lesssim$28\%, and $\lesssim$18\% average probability for hosting a double, triple, and quadruple system, respectively.

\subsubsection{Dusty vs. Clear Atmospheres}

\begin{figure}[t]
\figurenum{6}
\epsscale{1.1}
\plotone{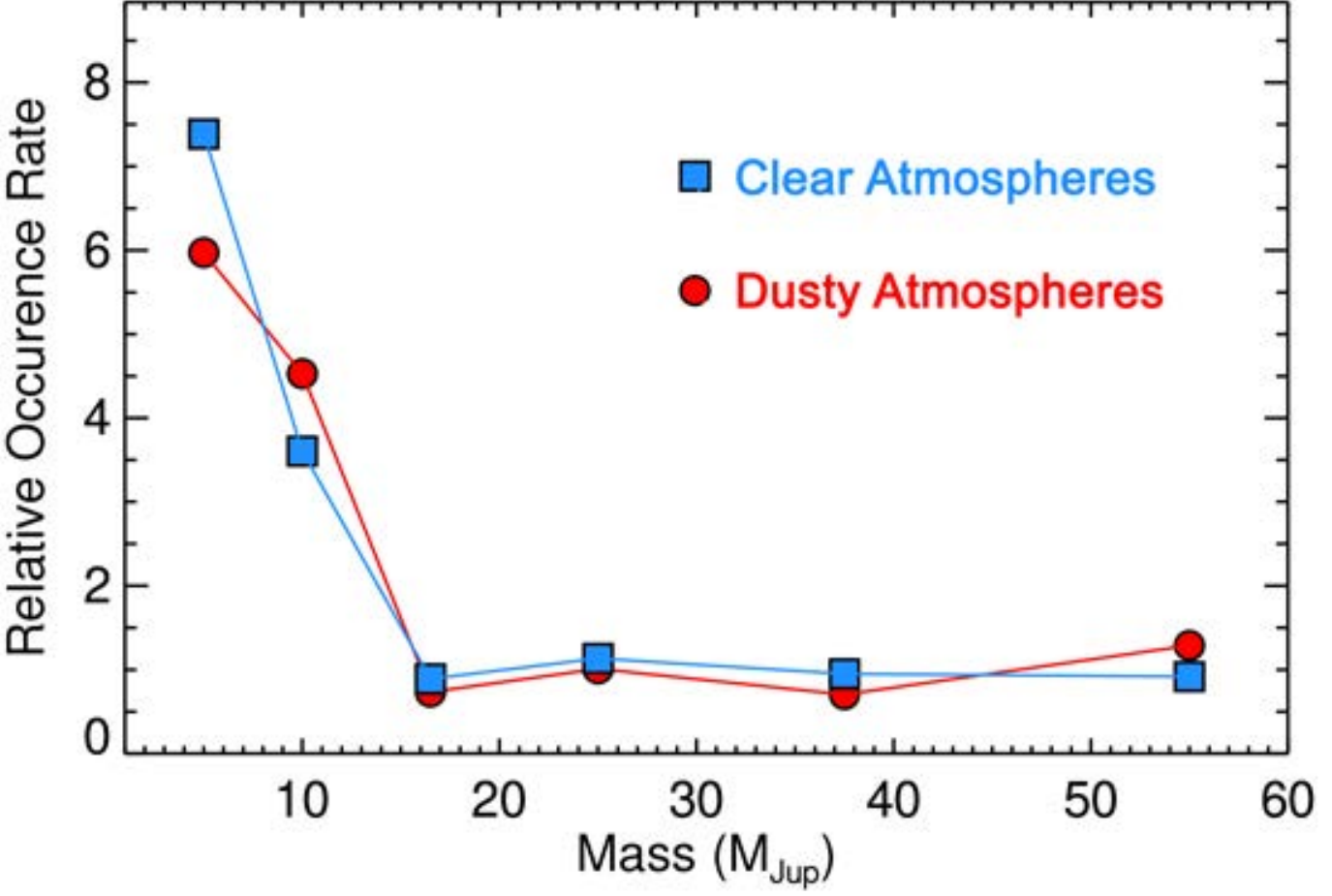}
\caption{The CMF assuming clear atmospheres (the \texttt{COND} grid) in blue, and dusty atmospheres (the \texttt{DUSTY} grid) in red. Both models assume the hot-start conditions described in \S3.4.1. The good agreement illustrates that our results are independent of the assumption of atmospheric dust content (or clouds) in converting photometric measurements to object masses. The uncertainties are similar to those in Fig. 3 \& Fig. 4.}
\end{figure}

\begin{figure*}[htp]
\figurenum{7}
\epsscale{1.1}
\plotone{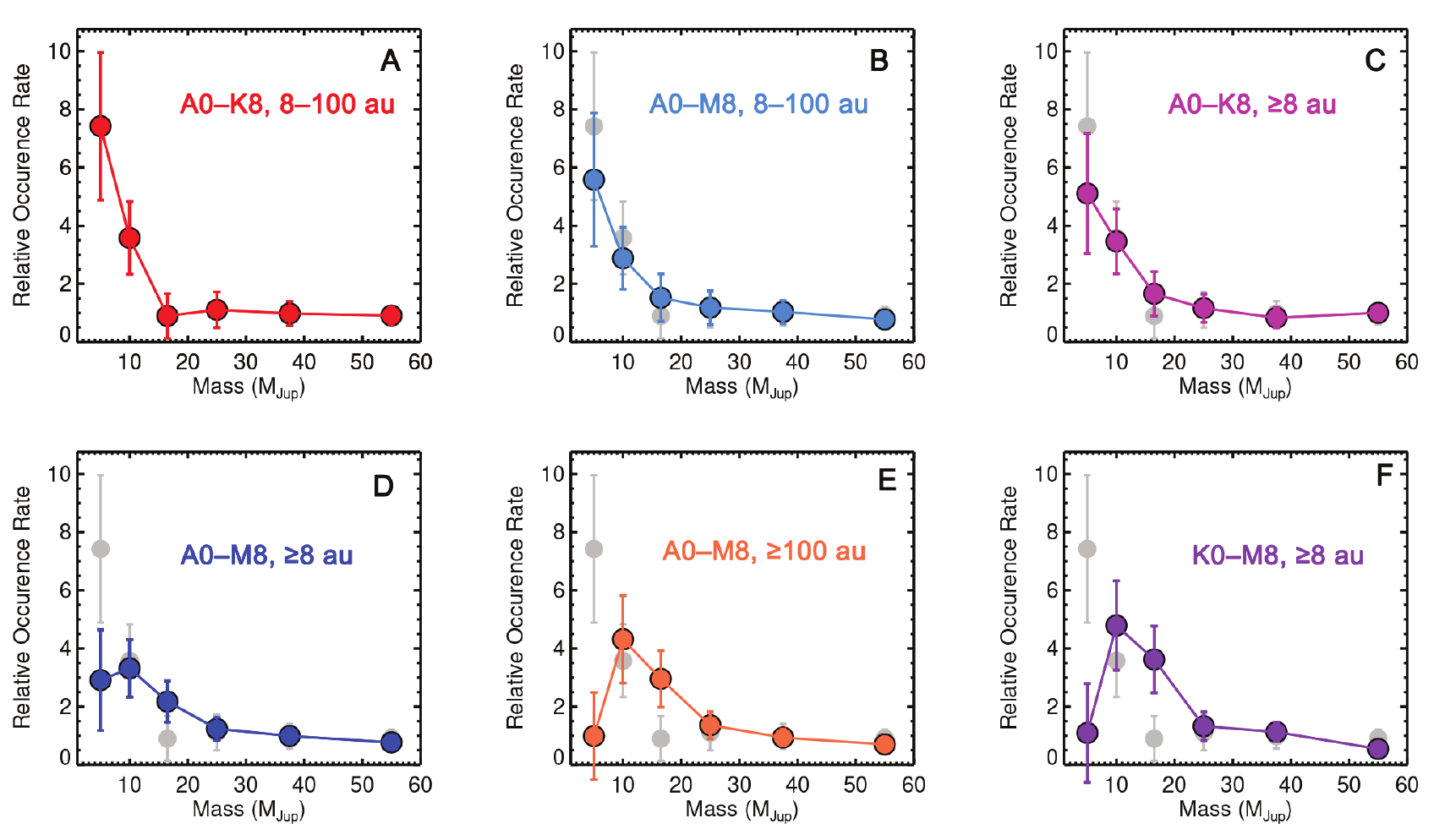}
\caption{The CMFs for various subsamples, assuming hot-start initial conditions and clear atmospheres. The primary sample (panel A) is represented in gray in panels B-F, and shows the steepest slope toward lower masses.}
\end{figure*}

A second assumption that may impact the results via conversion of photometry to mass is the choice of clear vs. dusty model atmospheres in the evolutionary grids. If an object has a significant fraction of dust (or clouds) in its photosphere, it will appear redder than the model predictions for clear atmospheres, and thus lead to a different interpretation of its mass. In the preceding sections, we utilized the clear model atmospheres of the \texttt{COND} grid \citep{Baraffe2003}. Now, we explore the effect of including dusty/cloudy model atmospheres of the \texttt{DUSTY} grid \citep{Chabrier2000} for applicable objects. We note that some objects, particularly T-dwarfs such as 51 Eri b \citep{Macintosh2015}, fail to be matched by this grid, which is more limited at lower masses and older ages. For these objects, we retain the mass distribution estimated from the \texttt{COND} grid, which is physically motivated by the fact that T-dwarfs (by definition) display primarily clear atmospheres. The results are shown in Fig. 6. We find a consistent result: the general trend is a CMF that is rising toward smaller masses, which verifies that the assumed dust content of the companions' photospheres does not significantly impact our results and conclusions.

\subsection{Exploration of Select Sub-samples}

In the preceding subsection, we have shown that the form of the CMF and limits on the fraction of systems hosting multiple wide-orbit giant companions are valid independent of model assumptions on initial planet luminosity and atmospheric dust content. However, the objects that we included in the preceding analysis were restricted to those within 100 AU of A0-K8 stars. While these choices were physically motivated, the effect of these assumptions deserves attention. For this purpose, we examine the effect of relaxing sample restrictions on spectral type, on orbital separation, and their combination. 

\begin{deluxetable}{ccccc}
\tabletypesize{\scriptsize}
\tablecaption{Sub-samples of Companions}
\tablewidth{0pt}
\tablehead{\colhead{ID} & \colhead{SpT}  & \colhead{Proj. Sep.}  & \colhead{\#} & \colhead{\S} }
\startdata
A & A0-K8 & 8$-$100 AU & 23 & 3.1-3.4 \\ 
B & A0-M8 &8$-$100 AU & 28 & 3.5\\ 
C & A0-K8 & $\geq$8 AU & 37 & 3.5\\ 
D & A0-M8 & $\geq$8 AU & 57 & 3.5\\ 
E & A0-M8 & $\geq$100 AU & 28 & 3.5\\ 
F & K0-M8 & $\geq$8 AU & 27 & 3.5\\ 

\enddata

\tablecomments{SpT: range of host spectral types, Proj. Sep: range of companion projected separations, \#: number of members, \S: relevant subsections in which the sub-sample is discussed. The lower limit of 8 AU in projected separation is imposed by the companion on the orbit with the shortest known period, $\beta$ Pictoris b.}
\end{deluxetable}

In addition, we explore several other subsamples, in pursuit of searching for potential differences in the CMF interior and exterior to 100 AU, and around hosts of different spectral type. These subsamples are listed in Table 2. We have so far focused on sub-sample A, which includes most of the objects typically considered to be directly imaged planets. This group has the added benefit of being partially isolated to effects of differing host mass and orbital configuration. Sub-samples B and C gradually relax these added selection criteria, by removing the spectral type criteria in B, and the projected separation criteria in C. Sub-sample D removes all selection criteria, and in other words includes all of the directly imaged substellar companions known to our study. Sub-sample E focuses on companions on \textit{very} wide-orbits ($\gtrsim$100 AU), and sub-sample F focuses on companions around late-type (and presumably low-mass) hosts. 




In our primary sample we did not include companions around M-stars. First, we explore the effect of including these companions (Sub-sample B). This search resulted in 28 companions, and is shown as the light blue line in Fig. 7, panel B. The result of including these additional companions produces only negligible effects on the companion mass distribution$-$namely a small decrease in the slope toward lower masses, although the effect is within the 1-$\sigma$ uncertainties. Instead, we also tried relaxing the separation criteria (Sub-sample C), but again exclude the companions around hosts of spectral type M. This search resulted in 37 companions, and is shown in the magenta line in Fig. 7, panel C. The distribution is still peaked toward planetary masses, with a $\sim$1$\sigma$ decrease in the frequency of the lowest mass bin with respect to the primary sample. Relaxing all selection criteria resulted in 57 companions (Sub-sample D), and is shown in the blue line in Fig. 7, panel D. While the distribution matches the others at higher masses, this sample shows a tentative peak at $\sim$10M$_{\rm Jup}$, while the frequency of the 3-7 M$_{\rm Jup}$ is consistent with the frequency of higher mass brown dwarfs. With respect to the primary sample, the relative frequency of the 3-7 M$_{\rm Jup}$ bin is reduced by $\sim$2$\sigma$. 

To explore the properties of very wide companions (wider than the 100 AU maximum considered earlier), we relax the selection criteria on spectral type, and focus on companions \textit{external} to 100 AU. This search resulted in 28 companions, and is represented in the orange line in Fig. 7, panel E. The behavior of the previous cases is enhanced, with the frequency of the lowest mass bin similarly reduced. This suggests that the similar behavior seen in the previous two samples is reflective of this very wide-orbit population. In this case, the frequency of 13-20 $M_{\rm Jup}$ is also enhanced by $\sim$2-$\sigma$ with respect to the CMF of the primary sample. In the final sub-sample, we examine K and M-type hosts, to explore differences around very late-type stars. This search resulted in 27 companions, and is represented in the purple line in Fig. 7, panel F. This sample shows a CMF that is similar to the previous case, with an enhancement only in the 7-13 M$_{\rm Jup}$ and 13-20 M$_{\rm Jup}$ bins, and a similar reduction of frequency of the 3-7 M$_{\rm Jup}$ bin. 

\subsubsection{Comparison of the Considered Samples}

\begin{figure}[htpb]
\figurenum{8}
\epsscale{1.1}
\plotone{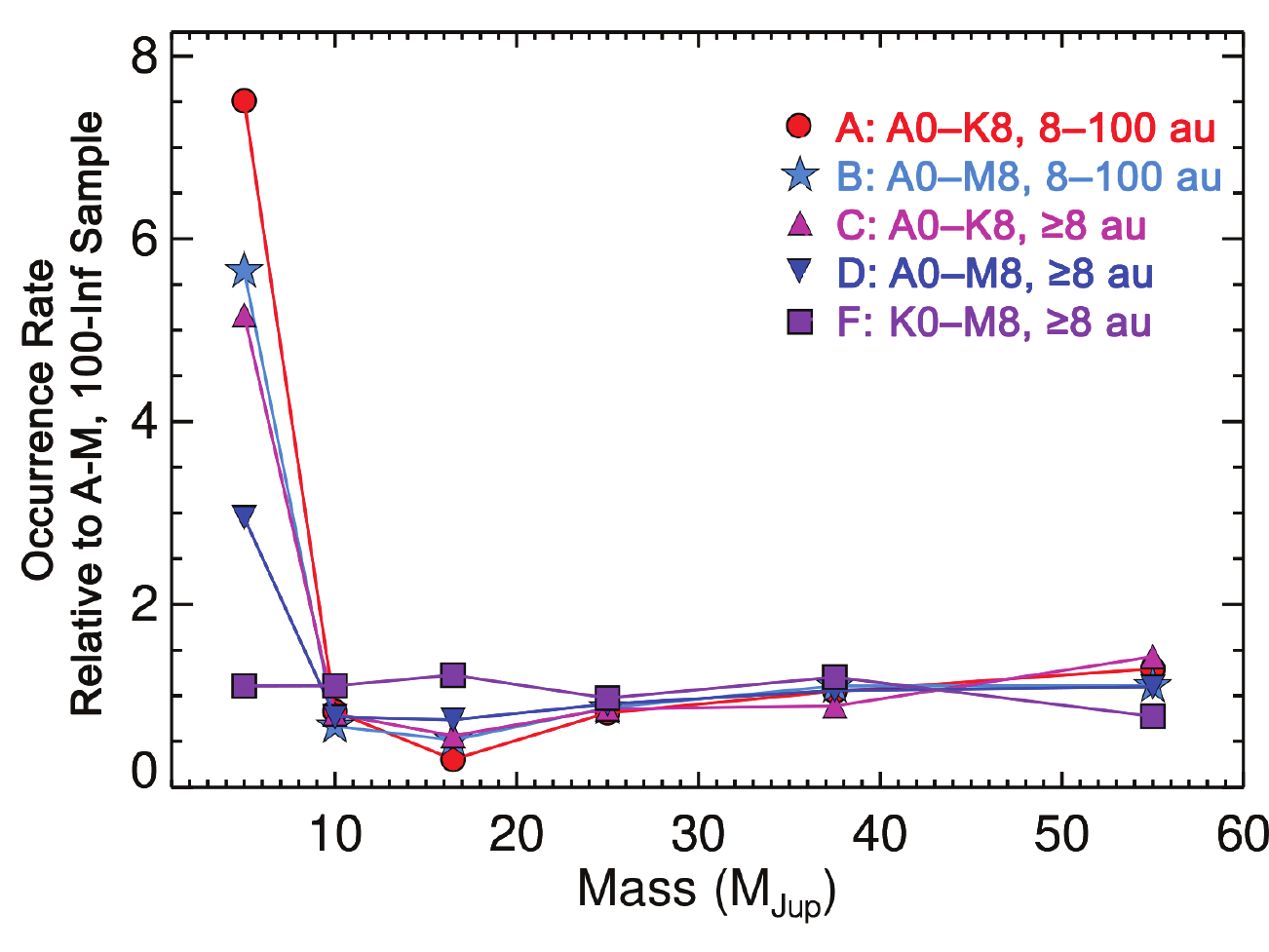}
\caption{The relative differences in the CMFs for various subsamples. Each CMF is re-normalized by the CMF of the A-M, 100-Inf sample. The most notable differences exist in the lowest mass bin, with subsamples restricted to within 100 AU showing a strong enhancement of 3-7 M$_{\rm Jup}$ planets compared to the other subsamples.}
\end{figure}

Now, we explore the differences between the CMFs for the various subsamples. To compare all of the subsamples simultaneously, it is convenient to re-normalize the samples compared to one that is selected as a ``standard" distribution. We choose to normalize by the A-M, $\geq$100 AU population: i.e., the very wide-orbit population, containing companions around any spectral type host (sub-sample E). This is a desirable choice, as it enables differences to be easily identified for populations containing a) companions on orbits consistent with typical sizes of protoplanetary disks ($\lesssim$100 AU), and b) companions around hosts of specific spectral types. The result is shown in Figure 8.


The most significant variation exists in the 3-7 M$_{\rm Jup}$ bin, with a strong a strong enhancement for planets of this mass among the samples restricted to within 100 AU. These show nearly an order of magnitude enhancement of 3-7 M$_{\rm Jup}$ planets compared to the other samples. Similarly, there is also an enhancement of the 3-7 $M_{\rm Jup}$ planets around early spectral type (A-K) hosts when orbital restrictions are relaxed. These effects are possibly further revealing of the formation mechanisms responsible for the various subsamples, and for the ensemble, which will be discussed in the next section.  

\section{Discussion}

The primary aim of our study was to investigate the relative mass function of giant planets and brown dwarfs (the planetary mass function, or CMF). The second aim of our study was to utilize this CMF to assess the probability that each system may in fact host multiple companions drawn independently from the same mass distribution. Here, we discuss those results in the context of predictions from various mechanisms that could have formed this population, and in the context of the results of other exoplanet surveys.

In \S3.1 we presented the CMF, which from the detected objects alone reveals a higher relative frequency of lower mass objects. The distribution becomes even more bottom heavy when incorporating information in the mass detection limits (\S2.2). In \S3.2 we compared the observed CMF to predictions from population syntheses, and to the observed CMF of the RV planets, and found a good agreement with both the CA predicted CMF and that of the RV population. A similar population synthesis-derived CMF representative of the GI scenario (followed by tidal downsizing for the lowest mass objects) does not match the observed form of the CMF, as it predicts a much lower frequency of planetary-mass companions relative to brown dwarf companions than is actually observed.

In \S3.2 we showed that the systems among our primary sample have (on average) a 68.2\% probability of hosting an additional (typically undetected) giant planet ($\geq$2 M$_{\rm Jup}$) whose mass is drawn independently from the same CMF. This is also in line with the predictions of a high fraction of systems with multiple companions resulting from the CA formation scenario. These simple results point strongly toward a CA origin for the wide-orbit giant planets, as GI is expected to produce a relatively flat CMF (similar to the stellar IMF at low masses) and a corresponding low fraction of systems hosting multiple wide-orbit substellar companions

To illustrate the robustness of this conclusion, we now turn to a critical assessment of our approach, considering its handling of observational biases, and its limitations. Finally, we discuss the general applicability of our results, and bearing the similarity of the CMF for wide-orbit and close-in giant planets, we argue for a general form of the CMF for planets within 100 AU. 

\subsection{Observational Biases}


We attempted to account for the observational biases by utilizing the statistical methods of survival analysis (as described in \S2.2), which enables information contained within the detection limits to be incorporated into the derived CMF. In this analysis, we assumed that each detection limit corresponds to one non-detected object. This choice was motivated in part by simplicity, but also for physical reasons. Given that the shape of the CMF from the detections alone points strongly toward a bottom-heavy CMF, and  thus a CA origin, it is reasonable to speculate that a high fraction of these systems hosts one or more additional companions of similar mass (perhaps close to 50\% as in \citealt{Knutson2014}). In reality, the average number of additional giant companions among these systems is very likely greater than zero, and of order unity, but is difficult to constrain further at present. 

Given that the detection limits typically correspond to planetary masses, it is not likely that unaccounted for observational biases would alter the inferred bottom-heavy form of the CMF. By assuming that the detection limits correspond to \textit{any} companions that actually exist, or none at all, we still arrive at a bottom-heavy CMF. As discussed in \S3.1, the inferred CMF is increasingly bottom-heavy as we assume that the detection limits correspond to a larger number of  objects that actually exist, but whose masses are beneath the detection limits. In order to change this picture, many brown dwarf companions above the survey detection limits would have necessarily gone unreported, which is not likely given that these were targeted specifically by past surveys (e.g., \citealt{Metchev2009}, \citealt{Brandt2014}, \citealt{Galicher2016}, \citealt{Vigan2017}, \citealt{Stone2018}, \citealt{Nielsen2019}). 

\subsection{Limitations and Generality of the Results}

One potential limitation of our approach, which focuses on systems that host directly imaged giant planets and brown dwarfs, is that these results are only necessarily applicable to systems hosting such companions, which may themselves be exceptional. In other words, it is possible that the CMF among these systems differs from that of the average star. Here we present counter-arguments to this point, and suggest that these results are likely to be generally applicable.

As a first consideration, these results may not be generally applicable if the systems considered here are non-representative of the general population (aside from the obvious and potentially exceptional property of hosting wide-orbit planets and brown dwarfs). Indeed, most of these systems share some similar properties, such as age (preferably young systems), and proximity, which are characteristics required to detect low-mass companions. Unless we reside in a particularly special location in the galaxy, there is no reason to expect that the nearby systems are different from the general population, so we can likely disregard the property of proximity. 

Youth, however, may play a role in causing the planets observed within the systems considered here to be exceptional. In particular, planets may experience significant orbital migration early in their lives. This is unlikely to bias our general results given that the median age of the systems considered here is 30 Myr, which is $\sim 10\times$ older than the typical disk lifetime. Nevertheless, some of the systems are young enough that they may still experience significant migration effects. In the case of outward migration, unless the companions are often ejected, they would still appear in the wide-orbit population, and have been considered as targets of this study. On the other hand, in the case that planets migrate inward, we may expect the mass functions of the inner and outer giant planets to be similar, which is indeed what is observed. Thus, it appears that youth may also be ruled out as a property that may cause our results to not be generally valid.



Given the similarity in the CMF, we suggest that planets discovered by the direct imaging and RV surveys, constituting outer and inner planets, respectively, are drawn from the same distribution. This could be taken as evidence that the inner population forms first at wide-orbits and subsequently migrates inward (or vice versa), or that both populations formed \textit{in situ} via similar processes. 

However, the picture is somewhat different at higher masses. While the inner and outer CMFs are of an overall bottom-heavy form, there are discernible differences among the relative frequency of brown dwarfs. The primary difference between the inner and outer CMFs is that there is a significant excess of brown dwarfs among the wide-orbit, directly imaged population compared to the mass function derived from RV surveys, and to the CMF predicted from CA population syntheses. This may be explained by a scenario in which GI is active in a minority of systems, and only in the outer regions, with the result typically being a wide-orbit brown dwarf or low mass stellar companion (as predicted by \citealt{Kratter2010}, etc.). 

\subsection{A Turn Over in the Wide-Orbit CMF?}

The wide-orbit CMF is best described by a distribution that is increasing sharply toward lower masses, similar to the form of the CMF of close-in planets. However, microlensing and transit surveys suggest that this behavior does not extend to arbitrarily low masses, and that a most likely companion-to-host mass-ratio exists. \cite{Suzuki2016} examined 22 planetary microlensing events from the MOA survey and inferred that the CMF follows a broken power law form with a peak at mass ratios of $q \sim$10$^{-4}$ (between Earth and Neptune's mass for M-F stars). \cite{Udalski2018} confirmed this trend in the eight planetary-mass microlensing detections of the OGLE survey and suggested a peak at $q \sim$1.7$\times 10^{-4}$. 

\cite{Pascucci2018} performed a similar analysis for the \textit{Kepler} planets, and found a break occurring at 2-3$\times$10$^{-5}$ that is universal among spectral types M-F. This break occurs at a mass ratio that is a few times lower than that for the microlensing planets, which are typically on wider orbits ($\gtrsim$1 AU) compared to those discovered by \textit{Kepler}, suggesting that the location of the peak in the CMF may shift toward higher mass ratios with increasing orbital separation. While our sample consists exclusively of objects on wide-orbits ($\geq$8 AU), we do not resolve a peak in the CMF, which is likely due to the fact that we are limited to much higher mass ratios above $q \gtrsim 10^{-3}$. Given the similarity between the CMFs of the wide-orbit and close-in planets (e.g., \citealt{Malhotra2015}, \citealt{Pascucci2018}, \citealt{Fernandes2018}), we may speculate that a similar break exists at lower mass ratios for the wide-orbit population, although this remains to be confirmed.

We note that in subsamples D, E, and F, we do observe a tentative peak in the CMF at $\sim$10 M$_{\rm Jup}$, or $q\sim0.01-0.05$. This peak occurs at much higher mass ratios than the peak in the CMF inferred from transit and microlensing surveys, and is within the range of mass ratios representative of stellar binaries. Given that these subsamples consist of primarily objects on very wide-orbits ($\geq$100 AU) and/or around very late spectral types, this behavior is likely due to vastly different formation and evolution processes than those that give rise to the break at $q\sim$10$^{-4}$ for close-in planets.


\subsection{Identifying the Dominant Formation Mechanism as a Function of Mass}

These results highlight an important difficulty in establishing a formation-motivated definition of what constitutes a ``planet" at high-masses$-$namely, that it is impossible to completely determine how a given object has formed from knowledge of solely its mass. While not a complete determination, the fact that the CMF is a superposition of a CA-like distribution and a GI-like distribution enables us to assign a probability that an object of a given mass formed via one of these two mechanisms (similar to \citealt{Reggiani2016}). Most objects beneath $\lesssim$10-20 M$_{\rm Jup}$ are representative of a CMF that is rising steeply toward lower masses, and thus likely originated via a CA-like formation process within a protoplanetary disk. Above $\gtrsim$10-20 M$_{\rm Jup}$, the opposite is true: most objects are representative of a flat CMF, similar to predictions from disk-born GI and the stellar IMF at low masses (e.g., \citealt{Kroupa2001}), and thus were likely born in a manner incorporating a rapid initial hydrodynamic collapse akin to star formation. However, between 10-20 $M_{\rm Jup}$ there is apparently a similar likelihood of forming via either mechanism. 

This is similar to the findings of \cite{Schlaufman2018}, who examined close in ($\leq$0.1 AU) transiting planets and brown dwarfs with Doppler inferred masses and found that bodies $\lesssim$10 M$_{\rm Jup}$ preferentially orbit solar-type dwarf stars with enhanced metalicity, while the same is not true for higher mass companions. Our study followed a different approach: we examined the relative mass function of wide-orbit companions ($\gtrsim$8 AU), and compared this to predictions from population synthesis models. We considered the point at which the synthetic mass distributions of CA and GI formed objects\footnote{With the latter scaled to the higher mass brown dwarfs.} intersected to be the dividing line between planets and brown dwarfs. Despite following a vastly different approach, we found a similar result, which supports the notion that objects on either side of $\sim$10-20 M$_{\rm Jup}$ primarily form via distinct physical processes.

\subsection{Exploration of Sub-samples: Evidence for Different Populations of Companions}

One possibility with the tools that we have developed for this study is to explore the differences in the relative mass function across different subsamples, which we have described in \S3.5. We explored several subsamples to investigate the effect of our initial assumptions of not including M-stars, and not including companions exterior to 100 AU in our primary analysis. This was essentially a consistency check, to ensure that these assumptions did not significantly impact our main results, which we verified in \S3.5. 

Additionally, we explored several subsamples for the purpose of investigating whether the CMF may vary interior/exterior to 100 AU, and around stars of different spectral types. We found that, in general, samples of companions interior to 100 AU show a more bottom-heavy mass function than samples restricted to companions exterior to 100 AU, and to hybrid populations that do not discriminate based on orbital separation.\footnote{We note also that planets on $\gtrsim$100 AU orbits could be captured free-floating planets. For predictions regarding the characteristics of such a population, see \cite{Perets2012}.} In other words, giant planets appear more frequently at smaller separations, although our data are not sensitive enough in the inner regions to resolve a turn-over in the distribution. From transit and RV planets, \cite{Fernandes2018} found a peak in the relative frequency of giant planets as a function of orbital separation at $\sim$2 AU, which is 4$\times$ smaller than the separation of our closest-in companion, $\beta$ Pic b. This implies that the relative frequency of giant planets continues to increase interior to the inner working angles of current high-contrast imaging facilities.

Likewise, we found that earlier spectral type hosts tend to have more bottom-heavy mass functions than later spectral type hosts. This is in contrast to the rocky planets, which occur more frequently around lower mass stars (e.g., \citealt{Mulders2015}), but which the observations considered here are insensitive to. The relative lack of 3-7 M$_{\rm Jup}$ planets around late spectral type hosts may reflect the unavailability of gas for giant planet formation within the protoplanetary disks around such low mass stars at later ages. In that case, the population of 7-13 M$_{\rm Jup}$ and higher mass companions could possibly represent either the population of companions born early on in the minority of gravitationally unstable disks around low mass stars, and/or those born even earlier as the low-mass end of binary star formation from turbulent fragmentation.

\subsection{An Order-of-Magnitude Assessment of The Frequency of HR 8799bcde-like Systems}

A second possible avenue to explore following these results is the nature of an intriguing system among those considered here: HR 8799 (\citealt{Marois2008}, \citealt{Marois2010}). This system is remarkable among those with directly imaged planets as one of the few systems with (detected) multiple planets, and the only known system with four wide-orbit super-Jupiters. On a superficial level, this is consistent with expectations assuming that companions are drawn independently from the CMF representative of the CA scenario: each of its four known companions are of planetary mass, which is the most likely incarnation of a system with such a large number of companions given their higher relative frequency.\footnote{Planetary masses are also expected for multiple companions on wide-orbits within the same system considering the requirements for dynamical stability (\citealt{Fabrycky2010}).} Additionally, one may argue that such systems are likely not atypical given its close proximity to Earth ($\sim$40 pc), unless the density of such systems is for some reason higher than average at our present galactic position. 

We can make some very basic (order of magnitude) estimates on the frequency of such systems based on the results presented in this study. We must first make an additional assumption about the absolute occurrence rates of substellar objects, since we have presented only relative mass functions. This is easier at the high mass end, given the incompleteness of existing surveys at lower masses. Surveys sensitive to such companions (e.g., \citealt{Metchev2009}, \citealt{Galicher2016}, \citealt{Vigan2017}, \citealt{Stone2018}, \citealt{Nielsen2019}) have shown that their frequency is on the order of a few percent, while the CMF presented here suggests approximately an order of magnitude more 3-7 M$_{\rm Jup}$ planets, comparable to those in HR 8799. 

Taking a pessimistic frequency of systems hosting a high mass brown dwarf to be $\sim$1\%, then nearly $\sim$10\% of systems would host a 3-7 M$_{\rm Jup}$ planet. Converting this into a probability that a single system will form multiple planets of this mass is more complicated, but as a simple approximation we will assume that the masses of the planets are drawn independently from the CMF. Thus, we arrive at a frequency of order $\sim$10$^{-4}$ for HR 8799-like systems. 


Within the 10 pc volume around the sun, there are roughly 400 known stars. Current instrumentation could detect an HR 8799-like system (a young system with multiple wide-orbit super-Jupiters) at a distance of $\sim$100 pc. Assuming the same stellar density as in the solar neighborhood, this volume contains roughly 4$\times$10$^{5}$ stars. Thus, if the occurrence rate of HR 8799 is roughly one in ten thousand stars, then there should be approximately forty such systems within 100 pc of Earth (though not all of these will be young). In other words, it is reasonable to expect that such a planetary system should exist at its proximity to the sun. However, this reasoning has not yet taken into account the fraction of stars that are young, and hence around which we could discover an HR 8799-like system. 

Assuming that the overall star formation rate is approximately flat in the galaxy, then the age distribution of stars (neglecting spectral type evolution) is also approximately flat, and consequently most stars will be quite old. For example, assuming an age distribution of 1 Myr to 10 Gyr results in only 0.4\% that are 40 Myr or younger, which in turn results in $\sim$1,600 young stars within 100 pc, and a roughly 16\% chance of observing such a system. While not impossible, the chance of this occurrence is still small enough to cause us to reconsider our initial assumptions.


One possibility is that the density of young stars in the solar neighborhood may be higher than the galactic average. This is consistent with our position within the local bubble$-$a structure that is thought to represent multiple supernova explosions approximately 10-20 Mya and possibly related to the formation of the Gould belt 30-60 Mya (\citealt{Berg2002}). This same event likely triggered further star formation, leading to an increased density of nearby young systems. If the local density of young stars is a few times higher than the galactic average, then we are left with a probability close to 100\% of detecting an HR 8799-like system. 

Additionally, if the fraction of wide-orbit brown dwarf companions to nearby stars is closer to a few percent, instead of the pessimistic 1\% assumed previously, this could further raise the probability of detecting an HR 8799-like system. Finally, it is possible that our initial assumption that planet masses are drawn independently from the CMF is incorrect. If systems hosting one super-Jupiter are in fact more likely to host an additional wide-orbit giant planet, this would further raise the probability of detecting an HR 8799-like system at its observed proximity and age. The latter scenario is supported, but not confirmed, by the result that a large fraction of the stars in our primary sample could host a second (typically undetected) wide-orbit planet more massive than 2 M$_{\rm Jup}$.



This analysis is highly over-simplified, but is useful as a sanity check on our results, and as an estimated frequency of HR 8799-like systems. If the CMF that we uncovered predicted no chance of detecting a system consisting of four super-Jupiters, or if it instead predicted that we should have detected many such systems, this would be reason to doubt the results. As it stands, the sanity check is consistent with the observations. While HR 8799 is likely a rare outcome of the planet formation process (on the order of one such system per ten-thousand stars), it is reasonable that one such young system exists relatively close to the sun.

\section{Summary and Conclusions}


1) We computed the relative mass distribution for companions within 100 AU around BAFGK stars. We found a steep function with a rising slope toward lower masses, in line with predictions for a core accretion formed population, and in remarkable agreement with the CMF of the inner planets detected by RV surveys. 

2) We estimated the probability that each system may host multiple wide-orbit giant companions drawn this distribution, and found that, on average, the systems considered here have a $\lesssim$68.2\% probability of hosting at least one additional wide-orbit giant companion whose mass is $\geq$2 M$_{\rm Jup}$. 

3) We verified that the above results are valid independent of model assumptions on initial planet luminosity (i.e., hot vs. cold start initial conditions), atmospheric dust content, and sample selection criteria (companions with projected separations $\leq$100 AU, excluding M-stars).

4) We suggested that these results are consistent with a scenario in which CA is the primary mechanism at forming companions less massive than $\sim$10$-$20 $M_{Jup}$, and that GI is the primary mechanism at forming higher mass companions. 

5) We explored the CMF of select subsamples, and find an enhanced population of super-Jupiters interior to 100 AU and around early-type hosts. 

6) As a sanity check, we estimated the frequency that these results would imply for HR 8799-like systems, and calculate the probability of detecting such a young system hosting multiple super-Jupiters with its proximity to the sun. We found that while HR 8799 is likely rare ($\sim 10^{-4}$ occurrence rate), it is reasonable that one such system has been discovered.

7) Our analysis suggests that future deep observations  of these and other targets should uncover a greater number of directly imaged planets, as the relative frequency of planets increases rapidly with decreasing mass. 

\section{Acknowledgments}

The authors acknowledge their sincere thanks to Jordan M. Stone and Thayne Currie for their insightful comments on an earlier version of this manuscript. The results reported herein benefited from collaborations and/or information exchange within NASA's Nexus for Exoplanet System Science (NExSS) research coordination network sponsored by NASA's Science Mission Directorate. KRW is supported by the National Science Foundation Graduate Research Fellowship Program under Grant No. 2015209499. This research has made use of the NASA Exoplanet Archive, which is operated by the California Institute of Technology, under contract with the National Aeronautics and Space Administration under the Exoplanet Exploration Program.




\appendix

\section{A: Properties of Companions}

In this appendix we present the table of companions and their relevant properties (Table 3), along with histograms of projected separations and host spectral type Figure (A1).

\begin{figure*}[htpb]
\figurenum{A1}
\epsscale{1.0}
\plotone{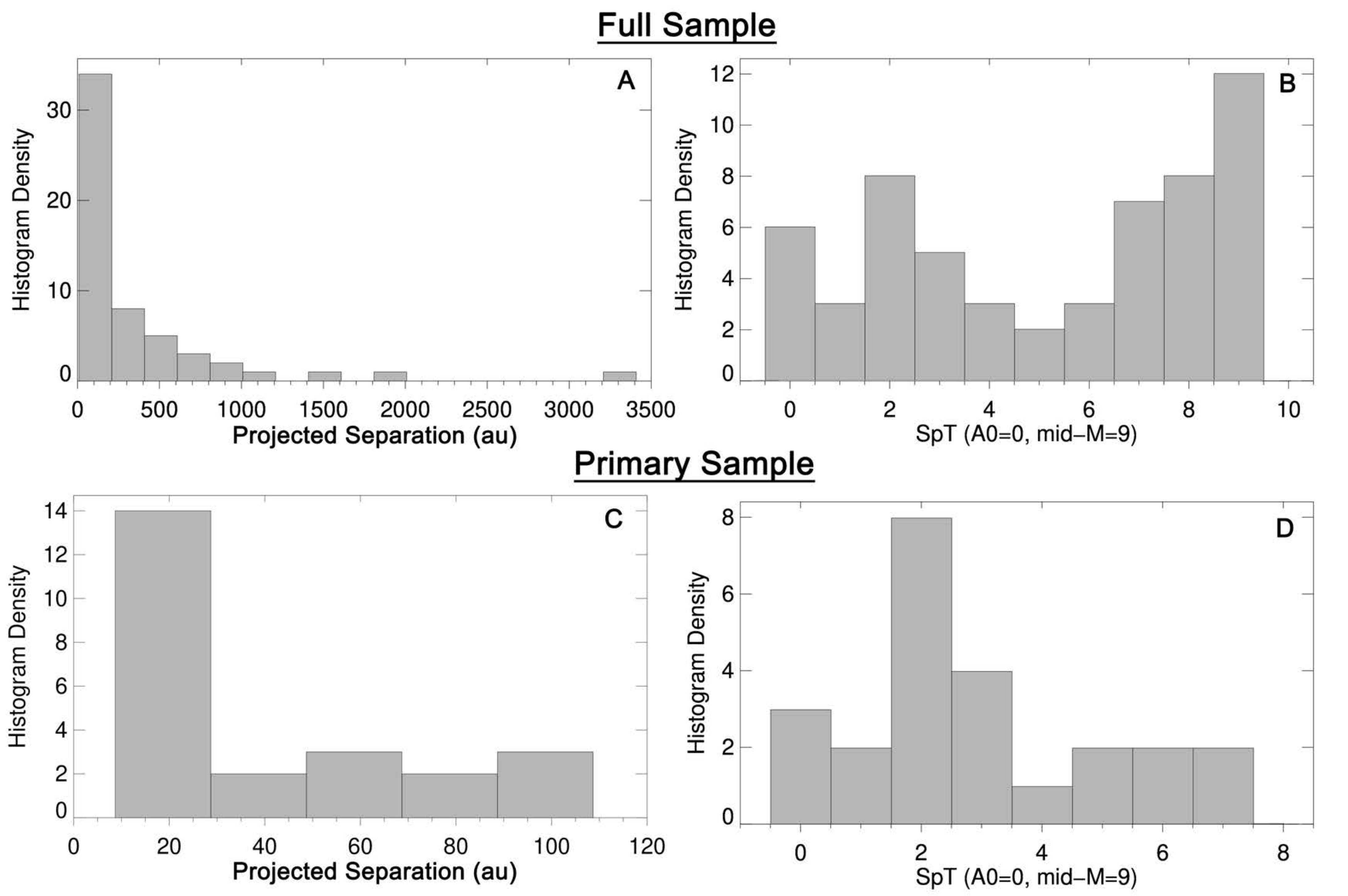}
\caption{Panel A: Histogram of companions' projected separation for the full sample. Panel B: Histogram of host spectral types for the full sample. Panel C: Histogram of companion's projected separation for the primary sample ($\leq$ 100 AU projected separation, host spectral type B-K). Panel D: Histogram of host spectral types for the primary sample. }
\end{figure*}

\LongTables

\begin{deluxetable*}{lcccccccccccc}
\tabletypesize{\footnotesize}
\tablecaption{Catalog of Directly Imaged Planets and Brown Dwarf Companions}
\tablewidth{0pt}
\tablehead{\colhead{Number} & \colhead{Name} & \colhead{Filter} & \colhead{App. Mag} & \colhead{$\Delta$mag} & \colhead{mag limit}  & \colhead{Dist. (pc)} & \colhead{$\Delta$Dist.} & \colhead{Age (Myr)} & \colhead{$\Delta$Age} & \colhead{Proj. Sep. (AU)} & \colhead{SpT}}
\startdata


{1} & {1RXS1609b} & {Ks} & {16.2} & {0.180} & {22.0} & {140.} & {1.30} & {5.00} & {2.00} & {307.} & {7.00} \\
{2} & {2M0103-55b} & {L$^{\prime}$} & {12.7} & {0.100} & {17.0} & {47.2} & {3.10} & {35.0} & {15.0} & {84.0} & {9.00} \\
{3} & {2M0122-24b} & {Ks} & {14.0} & {0.110} & {21.7} & {33.9} & {0.0860} & {120.} & {10.0} & {49.1} & {9.00} \\
{4} & {2M0219-39b} & {Ks} & {13.8} & {0.100} & {14.7} & {40.1} & {0.190} & {35.0} & {5.00} & {160.} & {9.00} \\
{5} & {2M2236+47b} & {Ks} & {17.4} & {0.0400} & {19.1*} & {69.7} & {0.160} & {120.} & {10.0} & {258.} & {7.00} \\
{6} & {2M2250+23b} & {Ks} & {14.9} & {0.0400} & {16.6} & {57.3} & {0.140} & {165.} & {35.0} & {510.} & {9.00} \\
{7} & {51Erib} & {Ks} & {17.5} & {0.140} & {19.5} & {29.8} & {0.120} & {20.0} & {6.00} & {13.4} & {2.00} \\
{8} & {ABPicb} & {Ks} & {14.1} & {0.0800} & {15.2*} & {50.1} & {0.0730} & {30.0} & {10.0} & {276.} & {9.00} \\
{9} & {CD-352722b} & {Ks} & {12.0} & {0.160} & {13.8*} & {22.4} & {0.0130} & {100.} & {50.0} & {67.2} & {8.00} \\
{10} & {CHXR73b} & {Ks} & {14.7} & {0.250} & {14.7*} & {191.} & {6.40} & {2.00} & {1.00} & {248.} & {9.00} \\
{11} & {CTChab} & {Ks} & {14.9} & {0.300} & {17.3} & {192.} & {0.770} & {2.00} & {1.00} & {518.} & {7.00} \\
{12} & {DHTaub} & {Ks} & {14.2} & {0.0200} & {16.8*} & {140.} & {20.0} & {2.00} & {2.00} & {322.} & {8.00} \\
{13} & {FWTaub} & {L$^{\prime}$} & {14.3} & {0.100} & {15.1*} & {140.} & {20.0} & {2.00} & {1.00} & {322.} & {9.00} \\
{14} & {GJ229b} & {Ks} & {14.6} & {0.100} & {15.5*} & {5.76} & {0.00151} & {1650} & {1350} & {44.9} & {8.00} \\
{15} & {GJ504b} & {L$^{\prime}$} & {16.7} & {0.170} & {17.4} & {17.5} & {0.0800} & {4000} & {1800} & {43.9} & {4.00} \\
{16} & {GJ570b} & {Ks} & {15.3} & {0.170} & {15.6*} & {5.88} & {0.00294} & {6000} & {4000} & {1520} & {7.00} \\
{17} & {GJ758b} & {L$^{\prime}$} & {16.0} & {0.190} & {16.0*} & {15.6} & {0.00540} & {4700} & {4000} & {28.1} & {6.00} \\
{18} & {GQLupb} & {L$^{\prime}$} & {11.7} & {0.100} & {12.6} & {152.} & {1.10} & {2.00} & {1.00} & {106.} & {7.00} \\
{19} & {GSC6214-21b} & {Ks} & {14.9} & {0.100} & {15.8*} & {109.} & {0.510} & {5.00} & {2.00} & {241.} & {8.00} \\
{20} & {GUPScb} & {Ks} & {17.7} & {0.0300} & {21.6} & {47.6} & {0.160} & {100.} & {30.0} & {2000} & {9.00} \\
{21} & {HD106906b} & {L$^{\prime}$} & {14.6} & {0.100} & {16.7} & {103.} & {0.400} & {13.0} & {2.00} & {734.} & {3.00} \\
{22} & {HD1160b} & {Ks} & {14.0} & {0.120} & {19.0} & {126.} & {1.20} & {165.} & {135.} & {96.9} & {0.00} \\
{23} & {HD19467b} & {Ks} & {18.0} & {0.0900} & {19.0*} & {32.0} & {0.0400} & {7300} & {2700} & {52.8} & {5.00} \\
{24} & {HD203030b} & {Ks} & {16.2} & {0.100} & {17.1*} & {39.3} & {0.100} & {90.0} & {60.0} & {468.} & {6.00} \\
{25} & {HD206893b} & {L$^{\prime}$} & {13.4} & {0.160} & {14.8} & {40.8} & {0.100} & {325.} & {275.} & {20.4} & {3.00} \\
{26} & {HD284149b} & {Ks} & {14.3} & {0.0400} & {19.1} & {118.} & {0.710} & {32.5} & {17.5} & {437.} & {4.00} \\
{27} & {HD4113c} & {Ks} & {19.7} & {0.120} & {20.4} & {41.9} & {0.0900} & {4800} & {1500} & {21.8} & {5.00} \\
{28} & {HD95086b} & {Ks} & {18.8} & {0.300} & {20.8} & {86.4} & {0.200} & {17.0} & {4.00} & {53.6} & {2.00} \\
{29} & {HD984b} & {Ks} & {12.2} & {0.0400} & {16.1} & {45.9} & {1.03} & {115.} & {85.0} & {9.18} & {3.00} \\
{30} & {HII1348b} & {Ks} & {14.9} & {0.100} & {15.7*} & {143.} & {0.996} & {113.} & {12.5} & {157.} & {7.00} \\
{31} & {HIP64892b} & {Ks} & {13.6} & {0.100} & {20.8} & {125.} & {1.40} & {20.0} & {11.0} & {150.} & {0.00} \\
{32} & {HIP65426b} & {Ks} & {16.6} & {0.300} & {18.8} & {109.} & {0.710} & {14.0} & {4.00} & {90.6} & {1.00} \\
{33} & {HIP73990b} & {L$^{\prime}$} & {13.3} & {0.400} & {13.8} & {111.} & {0.790} & {15.0} & {5.00} & {17.7} & {2.00} \\
{34} & {HIP73990c} & {L$^{\prime}$} & {13.2} & {0.450} & {13.8} & {111.} & {0.790} & {15.0} & {5.00} & {27.7} & {2.00} \\
{35} & {HIP74865b} & {L$^{\prime}$} & {12.8} & {0.300} & {14.3} & {124.} & {0.940} & {15.0} & {5.00} & {24.7} & {3.00} \\
{36} & {HIP77900b} & {Ks} & {14.0} & {0.01000} & {17.3*} & {151.} & {2.69} & {5.00} & {1.00} & {3300} & {9.00} \\
{37} & {HIP78530b} & {Ks} & {14.2} & {0.0400} & {15.9*} & {137.} & {1.50} & {5.00} & {1.00} & {618.} & {0.00} \\
{38} & {HNPegb} & {Ks} & {15.1} & {0.0300} & {17.2*} & {18.1} & {0.0200} & {350.} & {50.0} & {783.} & {4.00} \\
{39} & {HR2562b} & {Ks} & {16.6} & {0.140} & {20.1} & {33.6} & {0.300} & {475.} & {275.} & {20.2} & {3.00} \\
{40} & {HR3549b} & {Ks} & {15.1} & {0.100} & {22.0} & {95.4} & {0.810} & {125.} & {25.0} & {85.8} & {0.00} \\
{41} & {HR7329b} & {Ks} & {11.9} & {0.0600} & {13.2*} & {48.2} & {0.480} & {20.0} & {10.0} & {193.} & {0.00} \\
{42} & {HR8799b} & {L$^{\prime}$} & {15.7} & {0.120} & {18.7} & {41.3} & {0.100} & {30.0} & {10.0} & {70.9} & {2.00} \\
{43} & {HR8799c} & {L$^{\prime}$} & {14.8} & {0.0900} & {18.7} & {41.3} & {0.100} & {30.0} & {10.0} & {39.3} & {2.00} \\
{44} & {HR8799d} & {L$^{\prime}$} & {14.8} & {0.140} & {18.7} & {41.3} & {0.100} & {30.0} & {10.0} & {27.1} & {2.00} \\
{45} & {HR8799e} & {L$^{\prime}$} & {14.9} & {0.160} & {18.7} & {41.3} & {0.100} & {30.0} & {10.0} & {16.3} & {2.00} \\
{46} & {PDS70b} & {L$^{\prime}$} & {14.5} & {0.420} & {14.9} & {113.} & {0.520} & {5.40} & {1.00} & {22.1} & {7.00} \\
{47} & {PZTelb} & {Ks} & {11.5} & {0.0700} & {18.4} & {47.1} & {0.130} & {24.0} & {2.00} & {23.5} & {6.00} \\
{48} & {ROXs12b} & {L$^{\prime}$} & {12.6} & {0.0900} & {13.6*} & {137.} & {0.750} & {6.50} & {3.50} & {233.} & {8.00} \\
{49} & {ROXs42Bb} & {L$^{\prime}$} & {14.1} & {0.0900} & {15.3*} & {144.} & {1.50} & {2.00} & {1.00} & {159.} & {8.00} \\
{50} & {ROSS458ABc} & {Ks} & {16.5} & {0.0300} & {18.5*} & {11.5} & {0.0200} & {295.} & {145.} & {1170} & {8.00} \\
{51} & {SR12ABc} & {L$^{\prime}$} & {13.1} & {0.0800} & {14.2*} & {112.} & {5.17} & {5.25} & {4.75} & {977.} & {8.00} \\
{52} & {TWA5A(AB)b} & {L$^{\prime}$} & {12.1} & {0.100} & {12.9} & {49.4} & {0.140} & {10.0} & {5.00} & {98.8} & {9.00} \\
{53} & {Usco1610-19b} & {Ks} & {12.7} & {0.01000} & {18.2} & {144.} & {7.25} & {5.00} & {1.00} & {833.} & {9.00} \\
{54} & {Usco1612-18b} & {Ks} & {13.2} & {0.01000} & {18.2} & {158.} & {7.60} & {5.00} & {1.00} & {475.} & {9.00} \\
{55} & {BetaCirb} & {Ks} & {13.2} & {0.0400} & {16.5} & {28.4} & {0.330} & {435.} & {65.0} & {6200} & {1.00} \\
{56} & {BetaPicb} & {Ks} & {12.5} & {0.130} & {25.0} & {19.8} & {0.150} & {14.0} & {6.00} & {8.69} & {1.00} \\
{57} & {KappaAndb} & {L$^{\prime}$} & {13.1} & {0.0900} & {17.4} & {51.6} & {0.500} & {35.0} & {15.0} & {56.8} & {0.00} \\

\enddata
\tablecomments{Table A1:  Spectral Types: BA=0, A=1, AF=2, F=3, FG=4, G=5, GK=6, K=7, KM=8, M=9. \newline References: $^{1}$\cite{Lafreniere2008}, \cite{Lafreniere2010}, Gaia Collaboration Data Release 2 (GDR2);  $^{2}$ \cite{Delorme2013}, \cite{Riedel2014}, \cite{Zuckerman2000}, \cite{Torres2008}; $^{3}$ \cite{Bryan2016}, \cite{Bowler2013}, GDR2; $^{4}$ \cite{Artigau2015}, GDR2;  $^{5}$  \cite{Bowler2017}, GDR2;  $^{6}$ \cite{Desrochers2018}, GDR2; $^{7}$ \cite{Macintosh2015}, \cite{Samland2017}, GDR2;  $^{8}$ \cite{Chauvin2005}, GDR2; $^{9}$ \cite{Wahhaj2011}, GDR2; $^{10}$ \cite{Luhman2006}, GDR2; $^{11}$ \cite{Schmidt2009}, GDR2; $^{12}$ \cite{Itoh2005}; $^{13}$ \cite{Kraus2014}; $^{14}$ \cite{Nakajima1995}, \cite{Nakajima2015}, GDR2; $^{15}$ \cite{Kuzuhara2013}, \cite{Bonnefoy2018}, GDR2; $^{16}$ \cite{Burgasser2000}, GDR2; $^{17}$ \cite{Thalmann2009}, \cite{Currie2010}, \cite{Takeda2007}, GDR2; $^{18}$ \cite{Neuhauser2005}; $^{19}$ \cite{Ireland2011}, GDR2; $^{20}$ \cite{Naud2014}, GDR2; $^{21}$ \cite{Bailey2014}, GDR2; $^{22}$ \cite{Maire2016}, GDR2; $^{23}$ \cite{Crepp2014}, \cite{Crepp2015}, GDR2; $^{24}$ \cite{Metchev2006}, \cite{MilesPaez2017}, GDR2; $^{25}$ \cite{Delorme2013}, \cite{Milli2017}, GDR2; $^{26}$ \cite{Bonavita2017}, GDR2; $^{27}$ \cite{Cheetham2018a}, GDR2; $^{28}$ \cite{Rameau2013}, \cite{Galicher2014}, \cite{Chauvin2018}, GDR2; $^{29}$ \cite{Meshkat2015}, GDR2; $^{30}$ \cite{Geissler2012}, \cite{Meynet1993}, GDR2; $^{31}$ \cite{Cheetham2018b}, GDR2; $^{32}$ \cite{Chauvin2017}, GDR2; $^{33, 34, 35}$ \cite{Hinkley2015}, GDR2; $^{36}$ \cite{Aller2013}, GDR2; $^{37}$ \cite{Lafreniere2011}, GDR2; $^{38}$ \cite{Luhman2007}, \cite{Gaidos1998}, GDR2; $^{39}$ \cite{Mesa2016}, GDR2; $^{40}$ \cite{Mesa2018}, GDR2; $^{41}$ \cite{Lowrance2000}, GDR2; $^{42-45}$ \cite{Marois2008}, \cite{Marois2010}, \cite{Baines2012}, \cite{Maire2015}, GDR2; $^{46}$ \cite{Keppler2018}, \cite{Muller2018}, GDR2; $^{47}$ \cite{Maire2016}, GDR2; $^{48}$ \cite{Kraus2014}, GDR2; $^{49}$ \cite{Currie2014}, GDR2; $^{50}$ \cite{Burgasser2010}, GDR2; $^{51}$ \cite{Kuzuhara2011}, GDR2; $^{52}$ \cite{Lowrance1999}, GDR2; $^{53-54}$ \cite{Aller2013}, GDR2; $^{55}$ \cite{Smith2015}, GDR2; $^{56}$ \cite{Lagrange2010}, \cite{Currie2013}, GDR2, detection limit assumed from typical SPHERE and GPI performance; $^{57}$ \cite{Carson2013}, GDR2, detection limit from Stone et al. in prep (private communication). \newline The $Ks$ and $L^{\prime}$ labels denote the magnitude used for the model conversion of measured quantities to mass. Some observations were taken in a similar filter (e.g., broadband $K$ or $K1K2$), and in these cases we do not include the small ($\sim$0.1 mag) correction in magnitude, given that the introduced uncertainty is dwarfed by the typical measurement uncertainties. \newline * = estimated from the measurement uncertainties via $mag~limit = -2.5 \times log_{10}[5\times(10^{-\frac{App.~Mag}{2.5}}-10^{-\frac{App.~mag+\Delta mag}{2.5}}) ]$}

\end{deluxetable*}

\section{Appendix B:  Probability Distributions of Companion Masses and Detection Limits}

In Fig. B1 we present probability distributions for companion masses and upper mass limits on additional companions that are described in \S2.1. Many distributions show double-peaked and more complicated distributions, which is an artifact of degenerate mass-age-luminosity solutions within the measurement uncertainties. This effect is particularly prevalent for objects with large age uncertainties, as the spectral type transitions (representative of different cloud properties and atmospheric chemical abundances) result in non-linear evolution of photometric brightness with age \citep{Burrows2006}. 

\begin{figure*}[htpb]
\figurenum{B1}
\epsscale{1.1}
\plotone{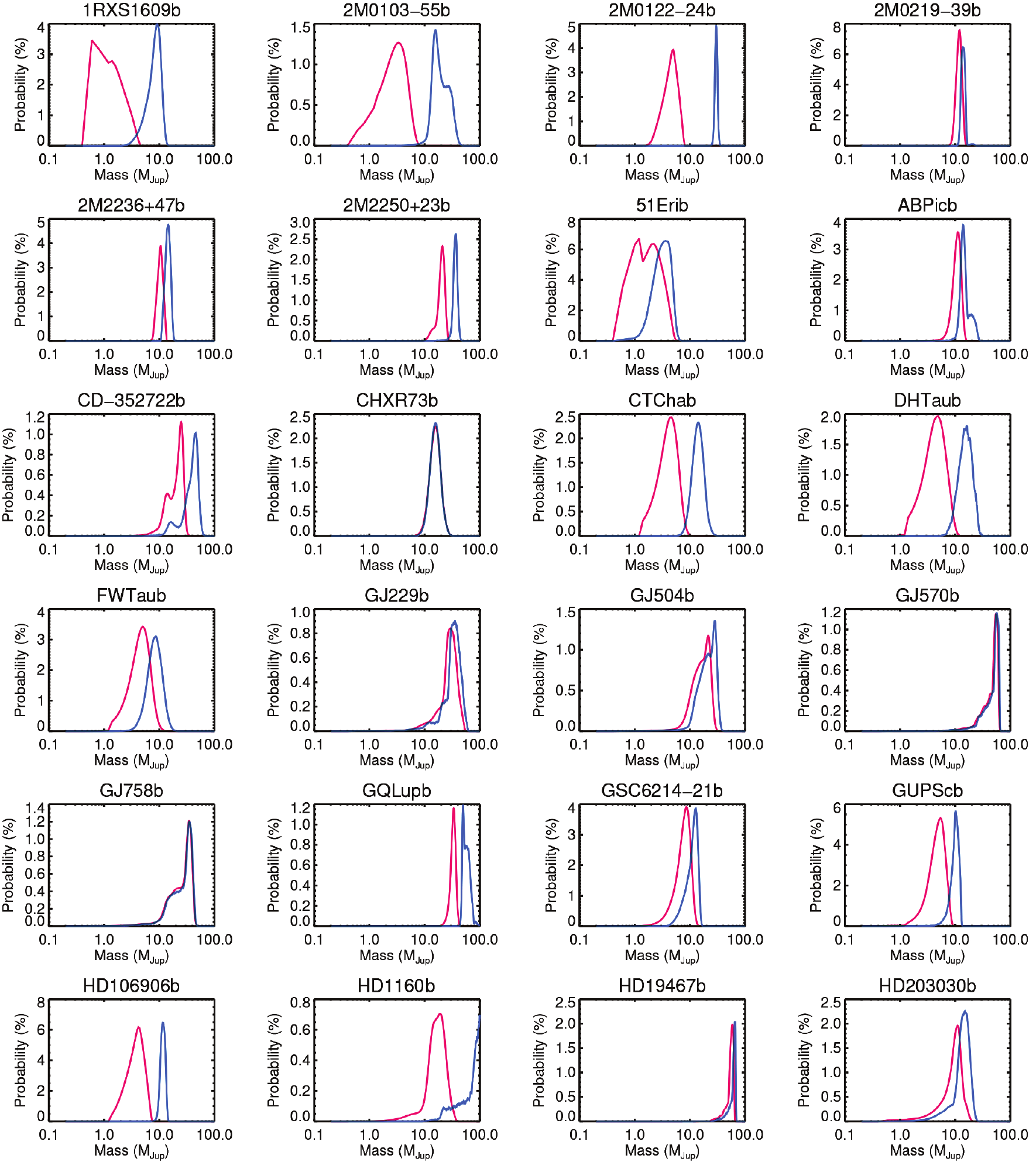}
\caption{Mass probability distributions (blue) and mass detection limit probability distributions (red) for all systems with directly imaged substellar companions considered in this study.}
\end{figure*}

\begin{figure*}[htpb]
\figurenum{B1}
\epsscale{1.1}
\plotone{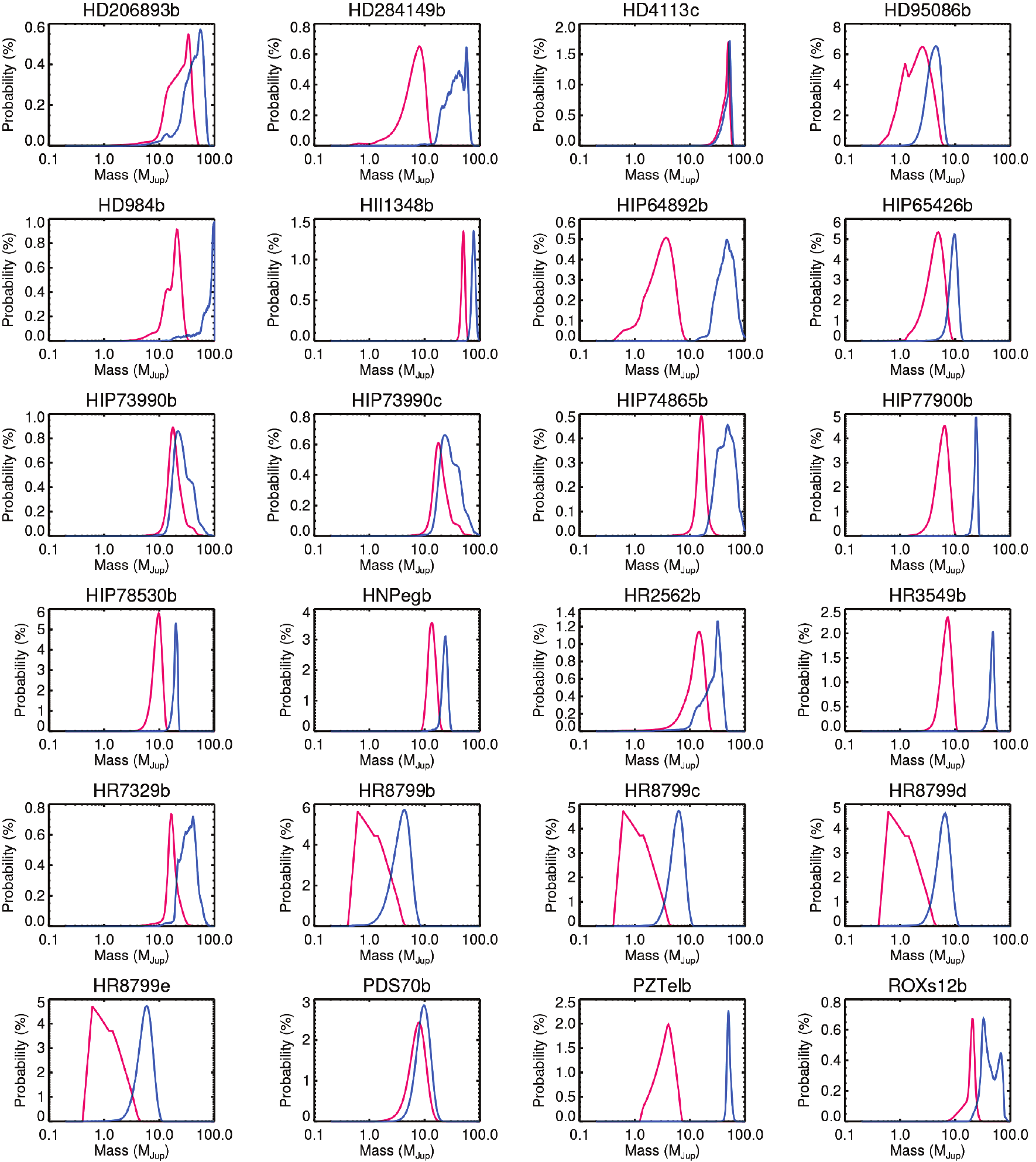}
\caption{Continued mass probability distributions (blue) and mass detection limit probability distributions (red).}
\end{figure*}

\begin{figure*}[htpb]
\figurenum{B1}
\epsscale{1.1}
\plotone{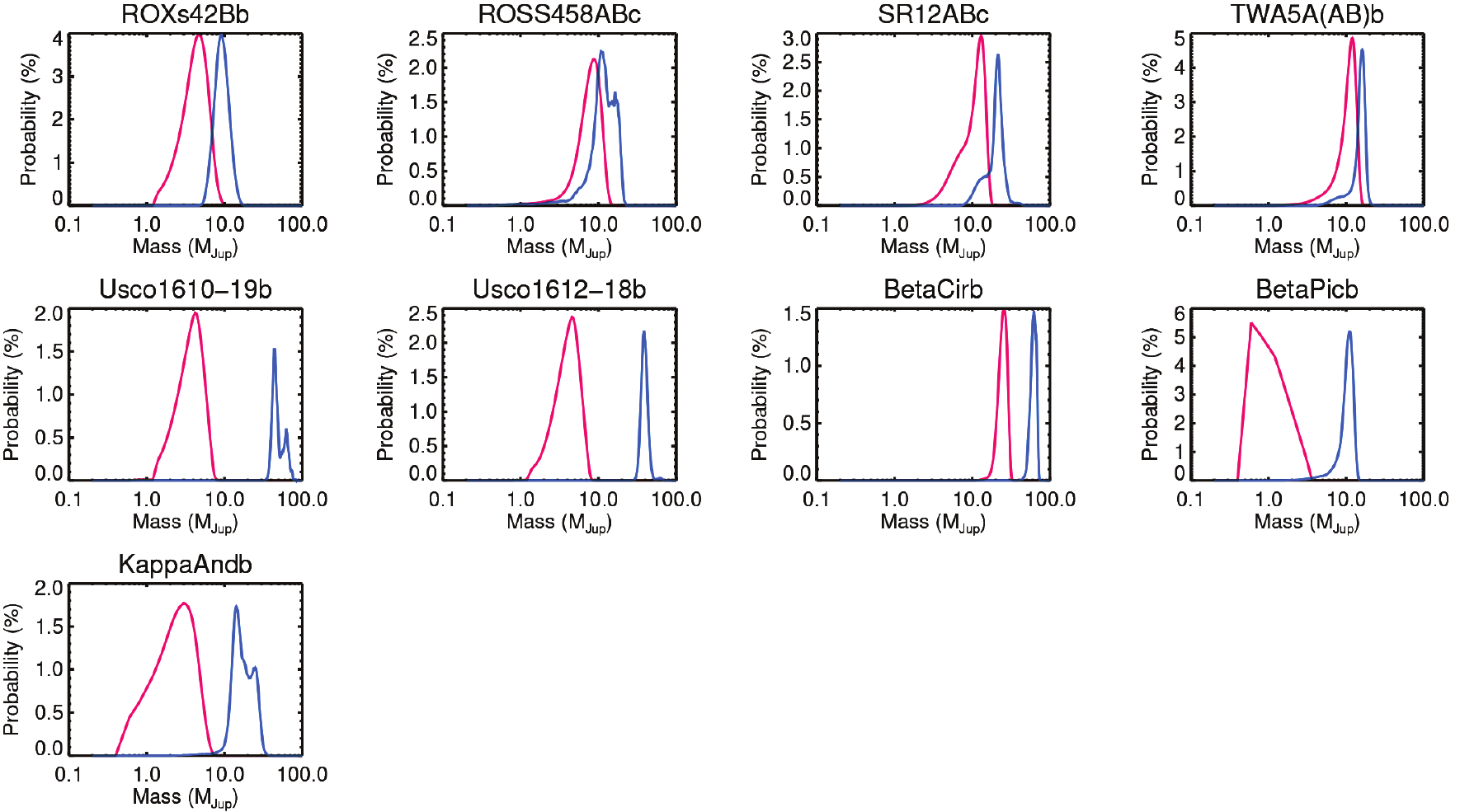}
\caption{Continued mass probability distributions (blue) and mass detection limit probability distributions (red).}
\end{figure*}

\clearpage






\end{document}